\documentclass[10pt]{article}

\usepackage{booktabs}
\usepackage{multirow}
\usepackage{comment}
\usepackage{amsmath,amssymb,epsfig,color,bbold}
\usepackage{feynmp-auto}
\usepackage[hidelinks]{hyperref}
\usepackage{caption}
\usepackage{subcaption}
\usepackage{cancel}
\usepackage{soul}
\usepackage{slashbox}
\usepackage{physics}
\usepackage{slashed}
\usepackage{psfrag}
\usepackage{graphicx,xspace}
\usepackage{bm}

\usepackage[affil-it, auth-sc]{authblk}
\usepackage{lineno}
\usepackage[sort&compress,numbers]{natbib}
\usepackage{doi}
\usepackage{eso-pic}
\usepackage[capitalise]{cleveref}

\newcommand{\be}{\begin{equation}}  
\newcommand{\ee}{\end{equation}} 
\def\slash#1{#1\!\!\!/\!\,\,}  
\newcommand{\nl}{\nonumber \\ }
\renewcommand{\order}{{\cal O}}
\long\def\symbolfootnote[#1]#2{\begingroup%
\def\thefootnote{\fnsymbol{footnote}}\footnote[#1]{#2}\endgroup}

\def\dd{\mathrm{d}} 
\def\e{\mathrm{e}}

\def\me{m}

\newcommand{\iu}{{\rm i}}

\allowdisplaybreaks

\begin{document}

\begin{fmffile}{fmfnotes} 
\fmfcmd{%
vardef middir(expr p,ang) = dir(angle direction length(p)/2 of p + ang) enddef;
style_def arrow_left expr p = shrink(.7); cfill(arrow p shifted(4thick*middir(p,90))); endshrink enddef;
style_def arrow_left_more expr p = shrink(.7); cfill(arrow p shifted(6thick*middir(p,90))); endshrink enddef;
style_def arrow_right expr p = shrink(.7); cfill(arrow p shifted(4thick*middir(p,-90))); endshrink enddef;}

\fmfset{wiggly_len}{3mm}
\fmfset{wiggly_slope}{75}
\fmfset{curly_len}{2mm}
\fmfset{zigzag_len}{2mm}

\fmfset{arrow_ang}{15}
\fmfset{arrow_len}{2.5mm}
\fmfset{decor_size}{3mm}

\fmfstraight

\AddToShipoutPictureFG*{\AtPageUpperLeft{\put(-60,-75){\makebox[\paperwidth][r]{FERMILAB-PUB-26-0829-T}}}}
\AddToShipoutPictureFG*{\AtPageUpperLeft{\put(-60,-60){\makebox[\paperwidth][r]{CERN-TH-2026-294}}}}

\title{\Large\bf
QED corrections to neutron and nuclear $\beta$-decay at order $\alpha^2$
}
\author[1]{Zehua~Cao}
\author[1,2]{Richard~J.~Hill}
\author[3]{Ryan~Plestid}
\author[4,1,2]{Peter~Vander~Griend}
\affil[1]{University of Kentucky, Department of Physics and Astronomy, Lexington, KY 40506 USA \vspace{1.2mm}}
\affil[2]{Fermilab, Theoretical Physics Department, Batavia, IL 60510, USA
\vspace{1.2mm}}
\affil[3]{Theoretical Physics Department, CERN, 1 Esplanade des Particules, CH-1211 Geneva 23, Switzerland\vspace{1.2mm}}
\affil[4]{Physics Department, Brookhaven National Laboratory, Upton, NY 11973, USA}

\date{\today}

\maketitle
\begin{abstract}

\noindent 
We compute two-loop QED corrections to neutron and nuclear beta decay, including full electron mass dependence and real radiation contributions. 
The result for the neutron,  
$\delta_{R,\,\rm static}= ( 29.199 \pm 0.001 )\times 10^{-3}$,  
implies a factor $\sim 70 $ reduction 
in the uncertainty of the 
low-energy radiative corrections compared to previous results obtained in 
the large energy/small electron-mass limit, $m_e/(m_n-m_p) \to 0$. For superallowed nuclear decays we find that the $Z$-independent order $\alpha^2$ corrections range from $(0.1-0.5) \times 10^{-4}$ i.e., of roughly the expected size and comparable to current uncertainty estimates. 
\end{abstract}

\newpage
\tableofcontents
\newpage

\section{Introduction}

The neutron and nuclear beta decay processes are critical for determining fundamental constants of Nature~\cite{Bopp:1986rt,Ando:2004rk,Darius:2017arh,Seng:2018yzq,Seng:2018qru,Fry:2018kvq,Czarnecki:2019mwq,Hayen:2020cxh,Seng:2020wjq,Gorchtein:2021fce,UCNt:2021pcg,Shiells:2020fqp,Markisch:2018ndu,UCNt:2025,Gorchtein:2023naa}. These processes currently dominate the determination of the quark mixing parameter $|V_{ud}|$ and 
probe potential physics beyond the Standard Model~\cite{Cirigliano:2013xha,Glick-Magid:2016rsv,Gonzalez-Alonso:2018omy,Glick-Magid:2021uwb,Falkowski:2021vdg,Brodeur:2023eul,Crivellin:2020ebi,Coutinho:2019aiy,Crivellin:2021njn,Crivellin:2020lzu,Cirigliano:2022yyo}. 
At current levels of experimental precision, control over theoretical predictions is required at the 100 ppm (i.e., $\sim 10^{-4}$) level, implying two-loop QED corrections and the systematic incorporation of 
enhanced corrections at higher loop order. 
Standard analyses combine the $O(\alpha)$ correction of Sirlin~\cite{Sirlin:1967zza} with the $O(Z\alpha^2)$ corrections~\cite{Sirlin:1986cc,Jaus:1986te} (recently superseded by Refs.~\cite{Cao:2025zxs,Crosas:2025xyv}) and estimates of higher-order effects~\cite{Czarnecki:2004cw,Marciano:2005ec,Hayen:2017pwg}. Effective field theory frameworks that organize these corrections systematically have been developed recently~\cite{Hill:2023acw,Hill:2023bfh,Cirigliano:2023fnz,Cirigliano:2024msg,Cirigliano:2024rfk,Cirigliano:2024nfi}.

In previous work~\cite{VanderGriend:2025mdc,Cao:2025lrw}, we computed 
two-loop virtual corrections to neutron beta decay.  By neglecting contributions suppressed by powers of $\me/E$, where $\me$ is the mass of the electron and $E$ is the energy of the electron, the result may be expressed in factorized form, in terms of universal soft and collinear factors, and a simplified hard function obtained in the massless electron limit.  
The largest uncertainty in the total low energy radiative correction, $\sim 0.07\times 10^{-3}$, 
is due to neglect of these power corrections. 

Here we compute the complete two-loop correction for arbitrary $\me/E$ reducing this error to negligible levels and providing, for the first time, a complete two-loop analysis of long-distance QED radiative corrections in both neutron and nuclear beta decays.
We update the total low-energy radiative correction, $\delta_{R,\rm static}$, for neutron beta decay, and tabulate corresponding corrections for nuclear transitions relevant to $|V_{ud}|$ determinations. 

The paper is organized as follows.
\Cref{sec:bkg} sets up the background-field decomposition of the heavy-light amplitude and the dispersion
relation that connects it to the imaginary part of the background-field diagrams, and illustrates both at one loop.
\Cref{sec:twoloop} presents the two-loop form factors, obtained by direct computation and, independently, from
the dispersion relation, together with the zero-recoil matching coefficient $E(m)$.
\Cref{sec:Zalpha2} extracts the $Z\alpha^2$ correction to nuclear beta decay with full electron-mass dependence,
and \cref{sec:HV} gives the two-loop virtual correction to the decay rate.
\Cref{sec:discussion} combines these with the real-virtual and double-real corrections to obtain
$\delta_{R,\rm static}$ for the neutron (\cref{tab:table1_fullF}) and the $Z$-independent $O(\alpha^2)$ corrections
for superallowed nuclear decays (\cref{tab:nuc}).
The appendices describe the evaluation of the two-loop integrals (\cref{app:integrals}), the real radiation
(\cref{app:real}), a numerical recipe for the electron spectrum (\cref{app:spectrum}), and the choice of
resummation scheme (\cref{app:resummation}).

\section{Background field analysis}\label{sec:bkg} 

Following Ref.~\cite{VanderGriend:2025mdc}, the scattering amplitude can be decomposed as
\begin{align}\label{eq:Mdiagrams}
    \vspace{6pt}
{\cal M}(-w-\iu 0) &\equiv 
\hspace{-10mm}
\parbox{35mm}{
\begin{fmfgraph*}(100,55)
  \fmfleftn{l}{3}
  \fmfrightn{r}{3}
  \fmfbottomn{b}{5}
  \fmf{phantom}{l2,v,r2}
  \fmffreeze
  \fmf{plain}{r3,x,v}
  \fmf{double}{r1,y,v}
  \fmffreeze
\fmfv{decor.shape=circle,decor.filled=shaded,decor.size=5mm}{v}
\end{fmfgraph*}
}
\quad 
=
\quad
\left.\left( \parbox{37mm}{
\begin{fmfgraph*}(100,55)
  \fmfleftn{l}{3}
  \fmfrightn{r}{3}
  \fmfbottomn{b}{5}
  \fmf{phantom}{l2,v,r2}
  \fmffreeze
  \fmf{plain}{r3,x,v}
  \fmf{double}{l2,y,v}
  \fmffreeze
\fmfv{decor.shape=circle,decor.filled=shaded,decor.size=5mm}{v}
\end{fmfgraph*} 
}\right)\right|_{v\to -v}
\nl
&=\quad
\parbox{30mm}{
\begin{fmfgraph*}(100,55)
  \fmfleftn{l}{3}
  \fmfrightn{r}{3}
  \fmfbottomn{b}{5}
  \fmf{phantom}{l2,v,r2}
  \fmffreeze
  \fmf{plain}{r3,x,v}
  \fmf{double}{l2,y,v}
  \fmffreeze
\fmfv{decor.shape=circle,decor.filled=shaded,decor.size=5mm}{v}
\end{fmfgraph*} 
}
\quad
+
\quad
\parbox{30mm}{
\begin{fmfgraph*}(100,50)
  \fmfleftn{l}{3}
  \fmfrightn{r}{3}
  \fmfbottomn{b}{5}
  \fmf{phantom}{l2,v,r2}
  \fmffreeze
  \fmf{plain}{r3,x,v}
  \fmf{double}{l2,v}
  \fmffreeze
  \fmf{photon}{x,b4}
  \fmfv{decor.shape=cross}{b4}
    \fmfv{decor.shape=circle,decor.filled=full,decor.size=2mm}{v}
\end{fmfgraph*}
}
\quad + \quad 
\left(\rule{0cm}{1.4cm}\right.
\parbox{35mm}{
\begin{fmfgraph*}(100,50)
  \fmfleftn{l}{3}
  \fmfrightn{r}{3}
  \fmfbottomn{b}{7}
  \fmf{phantom}{l2,v,r2}
  \fmf{phantom}{l1,b5,b6,r1}
  \fmffreeze
  \fmf{plain}{r3,x,y,v}
  \fmf{double}{l2,v}
  \fmffreeze
  \fmf{photon}{x,b6}
  \fmf{photon}{y,b5}
  \fmfv{decor.shape=cross}{b5}
  \fmfv{decor.shape=cross}{b6}
      \fmfv{decor.shape=circle,decor.filled=full,decor.size=2mm}{v}
\end{fmfgraph*}
}  
 \nonumber \\[5mm]
& 
\quad + \quad
\parbox{30mm}{
\begin{fmfgraph*}(100,50)
  \fmfleftn{l}{3}
  \fmfrightn{r}{3}
  \fmfbottomn{b}{7}
  \fmf{phantom}{l2,v,r2}
  \fmf{phantom}{l1,b5,b6,r1}
  \fmffreeze
  \fmf{plain}{r3,x,y,v}
  \fmf{double}{l2,z,v}
  \fmffreeze
  \fmf{photon}{x,b6}
  \fmf{photon,right}{y,z}
  \fmfv{decor.shape=cross}{b6}
      \fmfv{decor.shape=circle,decor.filled=full,decor.size=2mm}{v}
\end{fmfgraph*}
}
\quad + \quad 
\parbox{30mm}{
\begin{fmfgraph*}(100,50)
  \fmfleftn{l}{3}
  \fmfrightn{r}{3}
  \fmfbottomn{b}{7}
  \fmf{phantom}{l2,v,r2}
  \fmf{phantom}{l1,b5,b6,r1}
  \fmffreeze
  \fmf{plain}{r3,x,y,v}
  \fmf{double}{l2,z,v}
  \fmffreeze
  \fmf{photon,right}{x,z}
  \fmf{photon}{y,b5}
  \fmfv{decor.shape=cross}{b5}
      \fmfv{decor.shape=circle,decor.filled=full,decor.size=2mm}{v}
\end{fmfgraph*}
}
\quad + \quad 
\parbox{35mm}{
\begin{fmfgraph*}(100,50)
  \fmfleftn{l}{3}
  \fmfrightn{r}{3}
  \fmfbottomn{b}{5}
  \fmf{phantom}{l2,v1,r2}
  \fmffreeze
  \fmf{plain}{r3,x,v1}
  \fmf{double}{l2,v1}
  \fmffreeze
  \fmf{photon}{x,a1}
  \fmf{photon}{a2,b4}
  \fmf{fermion,left,tension=0.5}{a1,a2,a1}
  \fmfv{decor.shape=cross}{b4}
      \fmfv{decor.shape=circle,decor.filled=full,decor.size=2mm}{v1}
\end{fmfgraph*}
}
\\[6pt] \nonumber
& \quad + \quad
\parbox{35mm}{
\begin{fmfgraph*}(100,50)
\fmfstraight
  \fmfleftn{l}{3}
  \fmfrightn{r}{3}
  \fmfbottomn{b}{5}
  \fmf{phantom}{l2,v1,r2}
  \fmffreeze
  \fmf{plain}{r3,y,z,w,v1}
  \fmf{double}{l2,v1}
  \fmffreeze
  \fmf{photon}{z,b4}
  \fmf{photon,right}{y,w}
  \fmfv{decor.shape=cross}{b4}
      \fmfv{decor.shape=circle,decor.filled=full,decor.size=2mm}{v1}
\end{fmfgraph*}
}
\quad + \quad
\parbox{35mm}{
\begin{fmfgraph*}(100,50)
    \fmfstraight
  \fmfleftn{l}{3}
  \fmfrightn{r}{3}
  \fmfbottomn{b}{8}
  \fmf{phantom}{l2,v1,r2}
  \fmffreeze
  \fmf{plain}{r3,y,z,w,v1}
  \fmf{double}{l2,v1}
  \fmffreeze
  \fmf{photon}{y,b7}
  \fmf{photon,right=1.5}{z,w}
  \fmfv{decor.shape=cross}{b7}
      \fmfv{decor.shape=circle,decor.filled=full,decor.size=2mm}{v1}
\end{fmfgraph*}
} \quad 
\left.\rule{0cm}{1.4cm}\right)
+ \dots \,,\\[6pt]
\nonumber
\end{align}
where ${\cal M}(z)$ denotes the spacelike amplitude with four-velocities $v^\mu$ and $v^{\prime \mu}=p^\mu/m$, and $z=v\cdot v^\prime$.   
In Eq.~(\ref{eq:Mdiagrams}) and below: double lines denote a heavy charged particle; single solid lines denote the relativistic electron; wavy lines denote the photon; the cross denotes insertion of a background field~\cite{Borah:2024ghn}.
Blobs denote arbitrary loop diagrams with the given external particles. 

\subsection{Dispersion relation}

Since ${\cal M}(z)$ is analytic outside a branch cut extending from $z=-\infty \dots -1$, for any point outside this branch cut we have (after deforming the contour, and relabeling integration variable $z\to -z$)
\begin{align}\label{eq:dispersive}
    {\cal M}(w) &= \oint {dz\over 2 \pi \iu} {{\cal M}(z)\over z-w} 
    \nl
    &= {\cal M}(1) - (w-1)\int_{1}^{\infty} dz {1\over (z+w)(z+1)}{1\over 2\pi\iu}
    \big[ {\cal M}(-z-\iu 0) - {\cal M}(-z+\iu 0) \big] \,,
\end{align}
where we include a subtraction ${\cal M}(1)$ to ensure convergence at $|z|\to\infty$.  
Finally, we observe that 
\begin{align}
    {\cal M}(-z-\iu 0) - {\cal M}(-z+\iu 0)
    &= \big[ {\cal M}(-z-\iu 0) - {\cal M}(z-\iu 0) \big]
    - \big[ {\cal M}(-z+ \iu 0) - {\cal M}(z+\iu 0) \big]
\nl &\qquad 
    + \big[ {\cal M}(z-\iu 0) - {\cal M}(z+\iu 0)\big]
    \nl
    &= \big[ {\cal M}(-z-\iu 0) - {\cal M}(z-\iu 0) \big]
    - \big[ {\cal M}(-z+ \iu 0) - {\cal M}(z+\iu 0) \big]
   \nl
   &= 2\iu {\rm Im}\left[\rule{0cm}{1.4cm}\right.
    \!\!\! 
    \left.\left( \parbox{37mm}{
\begin{fmfgraph*}(100,55)
  \fmfleftn{l}{3}
  \fmfrightn{r}{3}
  \fmfbottomn{b}{5}
  \fmf{phantom}{l2,v,r2}
  \fmffreeze
  \fmf{plain}{r3,x,v}
  \fmf{double}{l2,y,v}
  \fmffreeze
\fmfv{decor.shape=circle,decor.filled=shaded,decor.size=5mm}{v}
\end{fmfgraph*} 
}\right)\right|_{v\to -v}
\!\!\!\!\!
-\quad
\parbox{37mm}{
\begin{fmfgraph*}(100,55)
  \fmfleftn{l}{3}
  \fmfrightn{r}{3}
  \fmfbottomn{b}{5}
  \fmf{phantom}{l2,v,r2}
  \fmffreeze
  \fmf{plain}{r3,x,v}
  \fmf{double}{l2,y,v}
  \fmffreeze
\fmfv{decor.shape=circle,decor.filled=shaded,decor.size=5mm}{v}
\end{fmfgraph*} 
}
\left.\rule{0cm}{1.4cm}\right] \,.
\end{align}
From Eq.~(\ref{eq:Mdiagrams}) we see that the required input to the dispersive integral 
involves all diagrams with at least one background field insertion. 

\subsection{One-loop}

Let us implement this program at one-loop order.   Recall the renormalized hard amplitude  is
\begin{align}
{\cal M}_H(w, \mu^{2}) = {\cal A}_H(w, \mu) + \frac{1}{w} \slash{v} {\cal B}_H(w, \mu) \,. 
\end{align}
To compute ${\cal A}_H$ and ${\cal B}_H$ using the dispersive formalism, we begin by noting that the relevant diagram is
\begin{align}\label{eq:bfdiags}
    \parbox{30mm}{
\begin{fmfgraph*}(100,50)
  \fmfleftn{l}{3}
  \fmfrightn{r}{3}
  \fmfbottomn{b}{5}
  \fmf{phantom}{l2,v,r2}
  \fmffreeze
  \fmf{fermion}{v,x,r3}
  \fmf{double}{l2,v}
  \fmffreeze
  \fmf{photon}{x,b4}
  \fmfv{decor.shape=cross}{b4}    \fmfv{decor.shape=circle,decor.filled=full,decor.size=2mm}{v}
\end{fmfgraph*}
}
~~~&= 
{\iu \alpha w\over \sqrt{w^2-1}}
\left( -4p^2 \over \mu^2 -\iu 0\right)^{-\epsilon}
\bigg[
{-1\over 2\epsilon} + {1\over 2(1-2\epsilon)}\left(-1 + {m\over E} \slash{v}\right)
\bigg]
\,,
\end{align}
from which we obtain, after renormalization and soft subtraction,  at renormalization scale $\mu=m$, 
\begin{align}\label{eq:imAB}
 {1\over \pi}{\rm Im}{\cal A}_{H\,{\rm 1bg}} &= {\alpha \over 2\pi}
 {w\over \sqrt{w^2-1}}\left( -1 + \ln[4(w^2-1)]\right)
 \,, \nl
 {1\over \pi}{\rm Im}{\cal B}_{H\,{\rm 1bg}} &= {\alpha \over 2\pi}
 {w\over \sqrt{w^2-1}} \,.
\end{align}
Substitution in Eq.~(\ref{eq:dispersive}), and using~\cite{Cao:2025lrw}
\begin{align}
     \lim_{z\to1} {\cal A}_H &= 1 + {\alpha \over 2\pi}(  -1
    ) \,,
    \nl
    \lim_{z\to1} {\cal B}_H &= {\alpha \over 2\pi} ( - 1 )\,,
\end{align}
we obtain the expected result
\begin{align}\label{eq:ah_bh}
        {\cal A}_H(z) &= 1 + {\alpha \over 2\pi}\bigg[
     zj(z) 
    - zJ(z) 
    \bigg] \,,
    \nl
    {\cal B}_H(z) &= {\alpha \over 2\pi}\left[ - z j(z) \right] \,.
\end{align}
The functions $j$ and $J$ are given for $w\ge 1$ by
\begin{align}
    wj(w) 
    &= \frac{w}{\sqrt{w^{2}-1}} \ln{(w+\sqrt{w^{2}-1})}
     \,, \nl
    wJ(w) 
    &= \frac{w}{\sqrt{w^{2}-1}} \left( {\rm Li}_{2}(1 - (w - \sqrt{w^{2}-1})^{2}) + \ln^{2}{(w+\sqrt{w^{2}-1})} \right) \,,
\end{align}
and the imaginary parts of the amplitudes (\ref{eq:imAB}) follow from the relations, with $\tilde{w} = -w-i0$, 
\begin{equation}
\begin{split}
    \tilde{w}j(\tilde{w}) =& wj(w) - \iu\pi \frac{w}{\sqrt{w^{2}-1}},\\
    \tilde{w}J(\tilde{w}) =& w J(w) - \iu\pi\frac{w}{\sqrt{w^{2}-1}} \ln(-4(w^{2}-1)-\iu0) \,.
\end{split}
\end{equation}

\section{Two-loop}\label{sec:twoloop}  

In this section we derive the two-loop form factor using two equivalent methods.  In the first method, we compute directly, at the timelike point ${\cal M}(-w-i0)$ appropriate for the final state proton in neutron beta decay.   In the second method, we use the dispersion relation to obtain the complete form factor (at either spacelike or timelike kinematics) from the imaginary part of background field diagrams, plus a subtraction point ${\cal M}(w=1)$.   
We note that the real part of the background field diagrams contribute to the $Z\alpha^2$ nuclear beta decay rate~\cite{Cao:2025zxs, Crosas:2025xyv}, and we quote results also for this component of the amplitude.  

\subsection{Direct computation}

We identify coefficients ${\cal A}_H$, ${\cal B}_H$ as matching coefficients for matching of a relativistic electron onto heavy-particle electron theory.  At $\overline{\rm MS}$ matching scale $\mu$ (here we do not separate the IR and UV scales, $\mu_S=\mu_H=\mu$), we write 
\begin{align}
    {\cal A}_H(w,\mu) &= 1 + {\bar{\alpha}\over 4\pi}A_H^{(1)} +
    \left({\bar{\alpha}\over 4\pi}\right)^2 A_H^{(2)} + \dots \,,
    \nl
     {\cal B}_H(w,\mu) &= \phantom{1 + }{\bar{\alpha}\over 4\pi}B_H^{(1)} +
    \left({\bar{\alpha}\over 4\pi}\right)^2 B_H^{(2)} + \dots \,,
\end{align}
where $\bar{\alpha}=\bar{\alpha}(\mu)$ is the  $\overline{\rm MS}$ QED coupling with a dynamical electron. 

For the spacelike amplitude (heavy anti-proton to electron), we have
the amputated diagrams, 
\begin{align}\label{eq:Mwdiagrams}
    \vspace{6pt}
{\cal M}(w)_{\rm bare} &\equiv 
\parbox{30mm}{
\begin{fmfgraph*}(100,55)
  \fmfleftn{l}{3}
  \fmfrightn{r}{3}
  \fmfbottomn{b}{5}
  \fmf{phantom}{l2,v,r2}
  \fmffreeze
  \fmf{plain}{r3,x,v}
  \fmf{double}{l2,y,v}
  \fmffreeze
\fmfv{decor.shape=circle,decor.filled=shaded,decor.size=5mm}{v}
\end{fmfgraph*} 
}
\nl
&=\quad
\parbox{30mm}{
\begin{fmfgraph*}(100,55)
  \fmfleftn{l}{3}
  \fmfrightn{r}{3}
  \fmfbottomn{b}{5}
  \fmf{phantom}{l2,v,r2}
  \fmffreeze
  \fmf{plain}{r3,x,v}
  \fmf{double}{l2,y,v}
  \fmffreeze
\fmfv{decor.shape=circle,decor.filled=full,decor.size=2mm}{v}
\end{fmfgraph*} 
}
\qquad
+ \quad 
\parbox{30mm}{
\begin{fmfgraph*}(100,50)
  \fmfleftn{l}{3}
  \fmfrightn{r}{3}
  \fmf{phantom}{l2,v,r2}
  \fmffreeze
  \fmf{plain}{r3,y,v}
  \fmf{double}{l2,z,v}
  \fmffreeze
  \fmf{photon,right}{y,z}
  \fmfv{decor.shape=circle,decor.filled=full,decor.size=2mm}{v}
\end{fmfgraph*}
}
\qquad
+ 
 \quad 
\parbox{30mm}{
\begin{fmfgraph*}(100,50)
  \fmfleftn{l}{3}
  \fmfrightn{r}{3}
  \fmf{phantom}{l2,v,r2}
  \fmffreeze
  \fmf{plain}{r3,x,y,v}
  \fmf{double}{l2,z,w,v}
  \fmffreeze
  \fmf{photon,right}{x,z}
  \fmf{photon,right}{y,w}
  \fmfv{decor.shape=circle,decor.filled=full,decor.size=2mm}{v}
\end{fmfgraph*}
}
\\[3mm]\nonumber
& \qquad + \quad
\parbox{30mm}{
\begin{fmfgraph*}(100,50)
  \fmfleftn{l}{3}
  \fmfrightn{r}{3}
  \fmf{phantom}{l2,v,r2}
  \fmffreeze
  \fmf{plain}{r3,x,y,v}
  \fmf{double}{l2,z,w,v}
  \fmffreeze
  \fmf{photon,right}{y,z}
  \fmf{photon,left}{x,w}
  \fmfv{decor.shape=circle,decor.filled=full,decor.size=2mm}{v}
\end{fmfgraph*}
}
\qquad
+ 
\quad
\parbox{30mm}{
\begin{fmfgraph*}(100,50)
  \fmfleftn{l}{3}
  \fmfrightn{r}{3}
  \fmf{phantom}{l2,v,r2}
  \fmffreeze
  \fmf{plain}{r3,x,y,w,v}
  \fmf{double}{l2,z,v}
  \fmffreeze
  \fmf{photon,right}{x,z}
  \fmf{photon,left=2}{y,w}
  \fmfv{decor.shape=circle,decor.filled=full,decor.size=2mm}{v}
\end{fmfgraph*}
}
\qquad
+ 
\quad
\parbox{30mm}{
\begin{fmfgraph*}(100,50)
  \fmfleftn{l}{3}
  \fmfrightn{r}{3}
  \fmf{phantom}{l2,v,r2}
  \fmffreeze
  \fmf{plain}{r3,x,v}
  \fmf{double}{l2,y,z,w,v}
  \fmffreeze
  \fmf{photon,right}{x,y}
  \fmf{photon,left=2}{w,z}
  \fmfv{decor.shape=circle,decor.filled=full,decor.size=2mm}{v}
\end{fmfgraph*}
}
\\[3mm]\nonumber
& \qquad + \quad
\parbox{30mm}{
\begin{fmfgraph*}(100,50)
  \fmfleftn{l}{3}
  \fmfrightn{r}{3}
  \fmf{phantom}{l2,v,r2}
  \fmffreeze
  \fmf{plain}{r3,x,y,w,v}
  \fmf{double}{l2,z,v}
  \fmffreeze
  \fmf{photon,right}{y,z}
  \fmf{photon,left}{x,w}
  \fmfv{decor.shape=circle,decor.filled=full,decor.size=2mm}{v}
\end{fmfgraph*}
}
\qquad
+ 
\quad
\parbox{30mm}{
\begin{fmfgraph*}(100,50)
  \fmfleftn{l}{3}
  \fmfrightn{r}{3}
  \fmf{phantom}{l2,v,r2}
  \fmffreeze
  \fmf{plain}{r3,x,v}
  \fmf{double}{l2,y,z,w,v}
  \fmffreeze
  \fmf{photon,right}{x,z}
  \fmf{photon,left}{w,y}
  \fmfv{decor.shape=circle,decor.filled=full,decor.size=2mm}{v}
\end{fmfgraph*}
}
\qquad
+
\quad
 \parbox{30mm}{
\begin{fmfgraph*}(100,50)
  \fmfleftn{l}{3}
  \fmfrightn{r}{3}
  \fmf{phantom}{l2,v,r2}
  \fmffreeze
  \fmf{plain}{r3,x,v}
  \fmf{double}{l2,y,v}
  \fmffreeze
  \fmf{phantom,right,tag=1}{x,y}        
  \fmffreeze
  \fmfipath{p}
  \fmfiset{p}{vpath1(__x,__y)}
  \fmfipair{a,b}
  \fmfiset{a}{point .4length(p) of p}
  \fmfiset{b}{point .6length(p) of p}
  \fmfipath{c}
  \fmfiset{c}{fullcircle scaled length(b-a) rotated angle(b-a) shifted .5[a,b]}
   \fmfi{photon}{subpath (0,.4length(p)) of p}
  \fmfi{photon}{subpath (.6length(p),length(p)) of p}
   \fmfi{plain}{subpath (0,4) of c}
  \fmfi{plain}{subpath (4,8) of c}
  \fmfv{decor.shape=circle,decor.filled=full,decor.size=2mm}{v}
\end{fmfgraph*}
}
\qquad + \dots \,,
\end{align}
where the ellipsis denotes three-loop and higher-order corrections.  
The bare, amputated, amplitude is calculated in dimensional regularization with $d=4-2\epsilon$ dimensions, in terms of the bare mass and coupling parameters.   Including onshell wavefunction renormalization  and $\overline{\rm MS}$ operator renormalization, the renormalized hard function is 
\begin{align} \label{eq:MmuMbare}
    {\cal M}(\mu) = Z_S  Z_H^{-1} Z_\psi^\frac12 Z_h^\frac12 {\cal M}_{\rm bare} \,, 
\end{align}
where $Z_H$ is the operator anomalous dimension, $Z_\psi$ and $Z_h$ are on-shell wavefunction renormalization factors for the lepton and heavy particle, and $Z_S$ is a further soft operator renormalization.

Let us present the renormalized results.  For $w\ge 1$, let us define:
\begin{equation*}
s=\sqrt{w^2-1},~ x=w-s,~ \varphi=\ln(w+s)=-\ln x,~ \ell_-=\ln(1-x^2)=\ln(2sx),~ \ell_+=\ln(1+x^2)=\ln(2wx).
\end{equation*}
Writing $w j(w) = w_j =\frac{w}{s}\varphi$ and $w J(w) = w_J =\frac{w}{s}\bigl[\mathrm{Li}_2(1-x^2)+\varphi^2\bigr]$, the one-loop results including $\mu$ dependence are, with $L=\ln(\mu/m)$,
\begin{align} 
{\cal A}_H^{(1)} &= 2(w_j-w_J) + (4w_j-1)\,L = \frac{w}{s}\Bigl[2\varphi-2\varphi^2-4\ell_-\varphi+2\,\mathrm{Li}_2(x^2)-\frac{\pi^2}{3}+4L\varphi\Bigr]-L \,,
\nl
{\cal B}_H^{(1)} &= -2w_j . \label{eq:AB1}
\end{align}
At two loops, we have 
\begin{align} \label{eq:AB2}
{\cal A}_H^{(2)} &= \tfrac12\bigl({\cal A}_H^{(1)}\bigr)^2 + \pi^2\,\frac{333 w^{3} + 153 w^{2} - 277 w - 113}{18 \left(w - 1\right) \left(w + 1\right)^{2}} + \frac{w}{s}\,a_{1} + \frac{w}{s^3}\,a_{3} + \frac{w^2}{s^2}\,a_{2} + \frac{1}{s^2}\,a_{0} + \frac{1}{s^4}\,a_{4} , \nl 
{\cal B}_H^{(2)} &= {\cal A}_H^{(1)}{\cal B}_H^{(1)} - \pi^2\,\frac{2 w \left(9 w^{2} + 3 w - 4\right)}{3 \left(w - 1\right) \left(w + 1\right)^{2}} + \frac{w}{s}\,b_{1} + \frac{w}{s^3}\,b_{3} + \frac{w^2}{s^2}\,b_{2} + \frac{1}{s^2}\,b_{0} + \frac{1}{s^4}\,b_{4} \,. 
\end{align}
Here $a_i$ and $b_i$ are functions of transcendental weight $\le 4$:
\begin{align}\label{eq:AB2vals}
a_{1}(\varphi) &= - \frac{4}{3} \varphi^{4} - \frac{52}{9} \varphi^{3} + \bigl(\frac{16 L}{3} + 16 \mathrm{Li}_{2}(-x^2) - 8 \mathrm{Li}_{2}(x^2) - 32 - \frac{8 \pi^{2}}{3}\bigr)\varphi^{2} \nonumber\\
  &\qquad  + \bigl(- \frac{32 L^{2}}{3} + L \left(\frac{32 \ell_{-}}{3} - \frac{16}{3}\right) - 64 \ell_{-} + 32 \mathrm{Li}_{3}(-x^2) - \frac{52 \pi^{2}}{9} - \frac{308}{27} + 24 \zeta\left(3\right)\bigr)\varphi \nl
  &\qquad  - \frac{16 L \mathrm{Li}_{2}(x^2)}{3} + 32 \mathrm{Li}_{2}(x^2) + 24 \mathrm{Li}_{4}(-x^2) + 12 \mathrm{Li}_{4}(x^2) \nl
  &\qquad  + \pi^{2} \left(\frac{8 L}{9} + 4 \mathrm{Li}_{2}(-x^2) + 16 \mathrm{Li}_{2}(x) - 8 \mathrm{Li}_{2}(x^2) - \frac{16}{3}\right) - \frac{9 \pi^{4}}{10} \nl
a_{3}(\varphi) &= - \frac{1}{3} \varphi^{4} - \frac{8}{3} \varphi^{3} + \bigl(- 2 \mathrm{Li}_{2}(x^2) - \pi^{2}\bigr)\varphi^{2} + \bigl(- 4 \mathrm{Li}_{3}(x^2) - \frac{8 \pi^{2}}{3} - \frac{32}{9} + 4 \zeta\left(3\right)\bigr)\varphi - 3 \mathrm{Li}_{4}(x^2) \nl
  &\qquad  + \pi^{2} \left(4 \mathrm{Li}_{2}(x) - 2 \mathrm{Li}_{2}(x^2)\right) - \frac{3 \pi^{4}}{10} \nl
a_{2}(\varphi) &= \frac{4}{3} \varphi^{4} + 4 \varphi^{3} + \bigl(- 32 \ell_{+} + 12 \ell_{-} + \frac{8 \pi^{2}}{3} + \frac{352}{9}\bigr)\varphi^{2}   + \bigl(32 \mathrm{Li}_{2}(-x^2) + 20 \mathrm{Li}_{2}(x^2) - \frac{2 \pi^{2}}{3}\bigr)\varphi + \frac{20 L^{2}}{3} \nl
  &\qquad  - \frac{35 L}{6} + 16 \mathrm{Li}_{3}(-x^2) + 26 \mathrm{Li}_{3}(x^2) + \pi^{2} \left(\frac{8 L}{3} - 8 \ell_{+} + 4 \ln(1+x) - 4 \ln{\left(2 \right)}\right) \nl
  &\qquad  - 18 \zeta\left(3\right) + \frac{11663}{432} \nl
a_{0}(\varphi) &= \frac{8}{3} \varphi^{3} + \bigl(8 \ell_{-} - \frac{110}{9}\bigr)\varphi^{2} + \bigl(- 8 \mathrm{Li}_{2}(x^2) + 8 \pi^{2}\bigr)\varphi \nl
  &\qquad  - \frac{20 L^{2}}{3} + \frac{35 L}{6} - 4 \mathrm{Li}_{3}(x^2) + \pi^{2} \left(- \frac{8 L}{3} + 16 \ln(1+x) - 4 \ln{\left(2 \right)}\right)  - \frac{10895}{432} + 2 \zeta\left(3\right) \nl
a_{4}(\varphi) &= \frac{16}{9} \varphi^{2}  \nl
b_{1}(\varphi) &= 4 \varphi^{3} + 16 \varphi^{2} + \bigl(\frac{16 L}{3} + 32 \ell_{-} + \frac{28}{9} + 4 \pi^{2}\bigr)\varphi - 16 \mathrm{Li}_{2}(x^2)  + \frac{8 \pi^{2}}{3} \nl
b_{3}(\varphi) &= \frac{1}{3} \varphi^{4} + \frac{8}{3} \varphi^{3} + \bigl(2 \mathrm{Li}_{2}(x^2) + \pi^{2}\bigr)\varphi^{2} + \bigl(4 \mathrm{Li}_{3}(x^2) - 4 \zeta\left(3\right) + \frac{16}{3} + \frac{8 \pi^{2}}{3}\bigr)\varphi \nl
  &\qquad  + 3 \mathrm{Li}_{4}(x^2) + \pi^{2} \left(- 4 \mathrm{Li}_{2}(x) + 2 \mathrm{Li}_{2}(x^2)\right) + \frac{3 \pi^{4}}{10} \nl
b_{2}(\varphi) &= -4 \varphi^{3} + \bigl(16 \ell_{+} - 12 \ell_{-} - \frac{32}{3}\bigr)\varphi^{2} + \bigl(- 16 \mathrm{Li}_{2}(-x^2) - 4 \mathrm{Li}_{2}(x^2) - \frac{14 \pi^{2}}{3}\bigr)\varphi \nl
  &\qquad  - 8 \mathrm{Li}_{3}(-x^2) - 10 \mathrm{Li}_{3}(x^2) + \pi^{2} \left(4 \ell_{+} - 12 \ln(1+x) + 4 \ln{\left(2 \right)}\right) - \frac{8}{3} + 10 \zeta\left(3\right) \nl
b_{0}(\varphi) &= - \frac{8}{3} \varphi^{2} \nl
b_{4}(\varphi) &= - \frac{8}{3} \varphi^{2}
\end{align}
We note that $b_0=b_4$ (so the ${\cal B}_H^{(2)}$ terms combine to $\frac{w^2}{s^4}b_4$), and $b_3=-a_3+\frac{16 \varphi}{9}$.

The form continues trivially to the timelike kinematics directly relevant for neutron beta decay. For $z>1$, ${\cal A}_H(-z-i0)$ and ${\cal B}_H(-z-i0)$
follow from Eq.~(\ref{eq:AB2}) with
\begin{equation*}
w\to -z,\quad s\to-\bar s,\quad x\to-\bar x,\quad \varphi\to\bar\varphi-i\pi,
\qquad \bar s=\sqrt{z^2-1},\ \bar x=z-\bar s,\ \bar\varphi=\ln(z+\bar s),
\end{equation*}
i.e.\ the prefactors $w/s$, $w/s^3$, $w^2/s^2$, $1/s^2$, $1/s^4$ and $\ell_\pm$, $\mathrm{Li}_n(\pm x^2)$ are unchanged,
$\mathrm{Li}_2(x)\to\mathrm{Li}_2(-\bar x)$, $\ln(1+x)\to\ln(1-\bar x)$, the rational $\pi^2$ terms take $w\to-z$, and only
$\varphi$ acquires $-i\pi$. Real parts follow by expanding the polynomials in $\varphi$.  For example, 
$\mathrm{Re}\,\frac12({\cal A}_H^{(1)})^2=\frac12[(\mathrm{Re}\,{\cal A}_H^{(1)})^2-(\mathrm{Im}\,{\cal A}_H^{(1)})^2]$ with the imaginary part being given by 
$\mathrm{Im}\,{\cal A}_H^{(1)}(-z-i0)=-2\pi\frac{z}{\bar s}\bigl(1-2\bar\varphi-2\ell_-+2L\bigr)$; this is evaluated with $\ell_-=\ln(1-\bar x^2)$.

\subsection{Dispersive computation} \label{sec:dispersive}

In this section we obtain the inputs to the dispersion relation (\ref{eq:dispersive}). 

\subsubsection{Zero recoil}

Let us begin by computing the subtraction point ${\cal M}(w=1)$, corresponding to the heavy-light form at zero recoil (finite-state electron at rest).  
From the results (\ref{eq:ah_bh}) and $\lim_{z\to 1} zj(z) = 1$, $\lim_{z\to 1} zJ(z) = 2$, we have at one loop, 
\begin{align}
    {\cal A}_H^{(1)}(w=1) &= -2 + 3 L \,, \nl
    {\cal B}_H^{(1)}(w=1) &= -2 \,, 
\end{align}
where $L=\ln(\mu/m)$.

At two-loop order, evaluation of our direct calculation at $w=1$ 
yields 
\begin{align}
    {\cal A}_H^{(2)}(w=1) &=
    \frac{2501}{144} + \frac{143\pi^2}{18} + 10\zeta_3 - \frac{52}{3}\pi^2\ln 2 + \frac{L^2}{2} + \Bigl(\frac{8\pi^2}{3} - \frac{13}{2}\Bigr)L
    \,, 
    \nl
{\cal B}_H^{(2)}(w=1) &= \frac{64}{9} + \frac{25\pi^2}{18} + 4\zeta_3 - \frac{8}{3}\pi^2\ln 2 - \frac{2}{3}L \,.
\end{align}

We note that the sum of ${\cal A}_H$ and ${\cal B}_H$ (using that for ${\cal B}_H$, we have $\slash{v} u(p) \to u(p)$ in the zero recoil limit), evaluated at $\mu=m$, is the QED limit of the result of Grozin~\cite{Grozin:2005tr}, which was extracted from the two-loop QCD results of Czarnecki and Melnikov~\cite{Czarnecki:1996gu,Czarnecki:1997cf,Franzkowski:1997vg}: 
\begin{align}\label{eq:Em}
    E(m) &= [{\cal A}_H + {\cal B}_H]_{w=1,\mu=m}
    = 1 + {\bar{\alpha}\over 4\pi} \left(-4 \right)
    + \left({\bar{\alpha}\over 4\pi}\right)^2 \bigg(  \frac{1175}{48} + \frac{28\pi^2}{3} + 14\zeta_3 - 20\pi^2\ln 2 \bigg) + \dots \,.
\end{align}
The quantity $E(m)$ appears in the two-step matching of $b\to c$ quark transitions onto HQET~\cite{Neubert:1992tg}.  
Our result represents a direct computation of this matching coefficient, to our knowledge for the first time.  

\subsubsection{Background field diagrams}

The background field diagrams (\ref{eq:Mdiagrams}) may be evaluated on an 
integral basis of heavy-particle propagators, using 
\begin{align}\label{eq:deltacut}
    {2\pi} \delta(v\cdot q) \propto {1\over v\cdot q + i0} + {1\over -v\cdot q + i0}
\end{align}
to represent the $\delta(v\cdot q)$ constraint.  

At one-loop,
\begin{align}
{\cal A}^{(1)}_{H, {\rm 1bg}} &= 
\frac{w}{s}\Bigl[2\pi^2+2\pi i\,(2\ell-1)\Bigr] \,,
\nl
{\cal B}^{(1)}_{H, {\rm 1bg}} &= 2\pi i\,\frac{w}{s} \,,
\end{align}
where $s=\sqrt{w^2-1}$, $x=w-s$, $\varphi=\ln(w+s)$, $L_s=\ln(2s)$, $L_w=\ln(2w)$, and
\begin{equation*}
\ell\equiv L_s-L=\ln\frac{2p}{\mu}\qquad(p=ms).
\end{equation*}
At two-loop order, there are contributions from one- and two-background field diagrams.  For two background fields: 
\begin{align}
\mathrm{Re}\,A^{(2)}_{H\,{\rm 2bg}} &= 16\pi^2\Bigl[\frac{w^2}{s^2}\Bigl(\frac{\pi^2}{24}-\frac{\ell^2}{2}+\frac{\ell}{2}\Bigr)+\frac54-\frac{\ell}{2}\Bigr], &
\mathrm{Im}\,A^{(2)}_{H\,{\rm 2bg}} &= 4\pi^3\Bigl[\frac{w^2}{s^2}(2\ell-1)+1\Bigr],  \label{eq:A2bg} \\
\mathrm{Re}\,B^{(2)}_{H\,{\rm 2bg}} &= -8\pi^2\,\frac{w^2}{s^2}\,\ell, &
\mathrm{Im}\,B^{(2)}_{H\,,{\rm 2bg}} &= 4\pi^3\,\frac{w^2}{s^2}. \label{eq:B2bg}
\end{align}
For one background field:
\begin{align}
\mathrm{Re}\,A^{(2)}_{H\,{\rm 1bg}} &=   \Bigl(- \frac{4 \pi^{2} w^{2}}{s^{2}} + \frac{2 \pi^{2} w \left(4 w^{2} - 3\right)}{s^{3}}\Bigr)\,\varphi^{2} - \frac{8 \pi^{2} w^{2}}{s^{2}}\,\varphi\,L_s + \frac{8 \pi^{2} w^{2}}{s^{2}}\,\varphi\,L \nonumber\\
&\quad + \Bigl(- \frac{8 \pi^{2} \left(3 w^{2} - 2\right)}{s^{2}} + \frac{4 \pi^{2} w \left(13 w^{2} - 7\right)}{3 s^{3}}\Bigr)\,\varphi \nonumber\\
  &\quad  - \frac{8 \pi^{2} \left(w^{2} + 4\right)}{s^{2}}\,L_s + \frac{32 \pi^{2} w^{2}}{s^{2}}\,L_w + \frac{8 \pi^{2} \left(w^{2} + 4\right)}{s^{2}}\,\ln(1-x) + \Bigl(- 8 \pi^{2} - \frac{22 \pi^{2} w}{3 s}\Bigr)\,L \nonumber\\
  &\quad  + \frac{8 \pi^{2} w \left(4 w^{2} - 3\right)}{s^{3}}\,\mathrm{Li}_2(-x) + \frac{4 \pi^{2} w^{2}}{s^{2}}\,\mathrm{Li}_2(x^2) - \frac{16 \pi^{2} w}{s}\,\mathrm{Li}_2(-x^2) \nonumber\\
  &\quad  - \frac{2 \pi^{2} \left(3 \pi^{2} w^{3} + 275 w^{3} + 3 \pi^{2} w^{2} + 185 w^{2} - 235 w - 153\right)}{9(w+1) s^{2}} \nonumber\\
  &\quad + \frac{2 \pi^{2} w \left(2 \pi^{2} w^{2} + 48 w^{2} - 48 - \pi^{2}\right)}{3 s^{3}} \label{eq:ReA1bg}
\end{align}
\begin{align}
\mathrm{Im}\,A^{(2)}_{H\,{\rm 1bg}} &=   \Bigl(- \frac{16 \pi w^{2}}{3 s^{2}} + \frac{4 \pi w \left(4 w^{2} - 3\right)}{3 s^{3}}\Bigr)\,\varphi^{3} + \frac{8 \pi w^{2}}{s^{2}}\,\varphi^{2}\,L_s - \frac{8 \pi w^{2}}{s^{2}}\,\varphi^{2}\,L - \frac{16 \pi w^{2}}{s^{2}}\,\varphi\,L_s^{2} + \frac{32 \pi w^{2}}{s^{2}}\,\varphi\,L_s\,L \nonumber\\
  &\quad  - \frac{16 \pi w^{2}}{s^{2}}\,\varphi\,L^{2} + \Bigl(- \frac{8 \pi \left(7 w^{2} - 1\right)}{s^{2}} + \frac{4 \pi w \left(13 w^{2} - 7\right)}{3 s^{3}}\Bigr)\,\varphi^{2} - \frac{8 \pi \left(w^{2} + 2\right)}{s^{2}}\,\varphi\,L_s + \frac{64 \pi w^{2}}{s^{2}}\,\varphi\,L_w \nonumber\\
  &\quad  + \frac{4 \pi w \left(4 w^{2} - 3\right)}{s^{3}}\,\varphi\,\mathrm{Li}_2(x^2) - \frac{32 \pi w}{s}\,\varphi\,\mathrm{Li}_2(-x^2) - \frac{44 \pi w}{3 s}\,L_s\,L + \frac{8 \pi w^{2}}{s^{2}}\,L_s\,\mathrm{Li}_2(x^2) + \frac{44 \pi w}{3 s}\,L^{2} \nonumber\\
  &\quad  - \frac{8 \pi w^{2}}{s^{2}}\,L\,\mathrm{Li}_2(x^2) + \Bigl(- \frac{4 \pi \left(185 w^{4} - 240 w^{2} + 63\right)}{9 s^{4}} + \frac{2 \pi^{3} w}{3 s^{3}}\Bigr)\,\varphi + \Bigl(- \frac{4 \pi^{3} w^{2}}{3 s^{2}} + \frac{64 \pi w}{s}\Bigr)\,L_s \nonumber\\
  &\quad  + \Bigl( \frac{4 \pi^{3} w^{2}}{3 s^{2}} + \frac{22 \pi w}{3 s}\Bigr)\,L - \frac{8 \pi \left(3 w^{2} - 1\right)}{s^{2}}\,\mathrm{Li}_2(x^2) + \frac{4 \pi w}{s^{3}}\,\mathrm{Li}_3(x^2) - \frac{32 \pi w^{2}}{s^{2}}\,\mathrm{Li}_2(-x^2) \nonumber\\
  &\quad  - \frac{32 \pi w}{s}\,\mathrm{Li}_3(-x^2) + \frac{4 \pi^{3}}{3} + \frac{- 4 \pi w \left(- 77 w^{2} + 162 w^{2} \zeta\left(3\right) - 135 \zeta\left(3\right) + 53\right)}{27 s^{3}}  - \frac{16 \pi w^{2}}{s^{2}}\,\varphi\,L \label{eq:ImA1bg}
\end{align}
\begin{align}
\mathrm{Re}\,B^{(2)}_{H\,{\rm 1bg}} &=  - \frac{2 \pi^{2} w}{s^{3}}\,\varphi^{2} - \frac{4 \pi^{2} w \left(3 w^{2} - 1\right)}{s^{3}}\,\varphi + \frac{24 \pi^{2} w^{2}}{s^{2}}\,L_s - \frac{16 \pi^{2} w^{2}}{s^{2}}\,L_w - \frac{24 \pi^{2} w^{2}}{s^{2}}\,\ln(1-x)  \nonumber\\
  &\quad  - \frac{8 \pi^{2} w}{s^{3}}\,\mathrm{Li}_2(-x)+ \frac{4 \pi^{2} w \left(11 w^{2} + 5 w - 4\right)}{3(w+1) s^{2}} - \frac{2 \pi^{2} w \left(24 w^{2} - 24 + \pi^{2}\right)}{3 s^{3}} \label{eq:ReB1bg}
\end{align}
\begin{align}
\mathrm{Im}\,B^{(2)}_{H\,{\rm 1bg}} &=  - \frac{4 \pi w}{3 s^{3}}\,\varphi^{3} + \Bigl( \frac{24 \pi w^{2}}{s^{2}} - \frac{4 \pi w \left(3 w^{2} - 1\right)}{s^{3}}\Bigr)\,\varphi^{2} + \frac{8 \pi w^{2}}{s^{2}}\,\varphi\,L_s - \frac{32 \pi w^{2}}{s^{2}}\,\varphi\,L_w + \frac{16 \pi w^{2}}{s^{2}}\,\varphi\,L  \nonumber\\
  &\quad - \frac{4 \pi w}{s^{3}}\,\varphi\,\mathrm{Li}_2(x^2) + \Bigl( \frac{8 \pi w^{2} \left(11 w^{2} - 9\right)}{3 s^{4}} - \frac{2 \pi^{3} w}{3 s^{3}}\Bigr)\,\varphi - \frac{32 \pi w}{s}\,L_s - \frac{22 \pi w}{3 s}\,L + \frac{8 \pi w^{2}}{s^{2}}\,\mathrm{Li}_2(x^2) \nonumber\\
  &\quad  - \frac{4 \pi w}{s^{3}}\,\mathrm{Li}_3(x^2) + \frac{16 \pi w^{2}}{s^{2}}\,\mathrm{Li}_2(-x^2) - \frac{4 \pi w \left(7 w^{2} - 9 \zeta\left(3\right) + 5\right)}{9 s^{3}} \label{eq:ImB1bg}
\end{align}

\subsubsection{Coulomb subtraction}

It is readily seen that the two-background field contribution to the imaginary part of the amplitude contains a non-integrable singularity at $z\to 1$.  Let us consider the subtractions in ${\cal M}(w)$.  Phrased in terms of the spacelike amplitude, 
\begin{align}
    {\cal A}^{(2)}_{H\,{\rm sub}}(w) &=  {-(4\pi)^2 \over 2(w+1)} \left[ -{\pi^2\over 12} - \frac12\left(\ln(w+1)+\ln{8} -L\right)^2 
    + \frac14\left(\ln(w+1)+\ln{8} -L\right) 
    \right] \,,
    \nl
    {\cal B}^{(2)}_{H\,{\rm sub}}(w) &=  {-(4\pi)^2 \over 2(w+1)} \left[
    -\frac14 \left(\ln(w+1)+\ln{8} -L\right) 
    \right] \,. \label{eq:sub}
\end{align}
Note that these subtractions are analytic outside $w\le -1$, and at $w\to -z-i0$
account for the $1/\beta^2$ singularity in ${\cal A}_{H,\rm 2bg}^{(2)}$, ${\cal B}_{H,\rm 2bg}^{(2)}$. 
The subtracted amplitudes, ${\cal A}_H^{(2)} \to {\cal A}_H^{(2)}- {\cal A}_{H,\rm sub}^{(2)}$, ${\cal B}_H^{(2)} \to {\cal B}_H^{(2)}- {\cal B}_{H,\rm sub}^{(2)}$, 
may be reconstructed from the dispersion relation, and are 
identical to the direct computation.  

\section{$Z\alpha^2$ correction to nuclear beta decay}\label{sec:Zalpha2} 

The real part of the two-loop background field diagrams yield the $Z\alpha^2$ correction to the nuclear beta decay rate~\cite{Cao:2025zxs,Crosas:2025xyv}.  
To compare with the conventions of Ref.~\cite{Cao:2025zxs},  
the one- and two-background photon terms above enter with factors 
$(-Z)$ and $(-Z)^2$.  Let us write (again using the $\overline{\rm MS}$ coupling)
\begin{equation*}
\frac{d\Gamma}{dE}=\Bigl(\frac{d\Gamma}{dE}\Bigr)_{\rm tree}\tilde F(Z\bar\alpha,E,\mu)\Bigl[1+\frac{\bar\alpha}{2\pi}\Bigl(\hat g+\frac{11}{4}\Bigr)+Z\bar\alpha^2\,K(w,L)+\dots\Bigr] \,.
\end{equation*}
The quantities ${\cal A}_{H\,\rm 1bg}^{(1)}$, ${\cal B}_{H\,\rm 1bg}^{(1)}$, 
${\cal A}_{H,\rm 2bg}^{(2)}$ and ${\cal B}_{H,\rm 2bg}^{(2)}$ reproduce the 
$Z\bar{\alpha}$ and $Z^2\bar{\alpha}^2$ terms of the $\overline{\rm MS}$ Fermi function $\tilde{F}$ \cite{Hill:2023acw,Hill:2023bfh}, which is given to all-orders in $Z\bar{\alpha}$ by, 
\begin{equation}    \label{FF-def}
   \tilde{F}(Z \bar{\alpha} ,E,\mu) =  \frac{4 \eta}{(1+\eta)^2} \frac{2 (1+\eta)}{[\Gamma(1+2\eta)]^2} |\Gamma(\eta +\iu\xi)|^2 \e^{\pi \xi} (2 p \e^{-\gamma_{\rm E}}/\mu)^{2(\eta-1)}~,
\end{equation}
where $\eta=\sqrt{1-(Z\bar{\alpha})^2}$ and $\xi= Z\bar{\alpha}/\beta$. 
The quantities ${\cal A}_{H}^{(1)}$ and ${\cal B}_{H}^{(1)}$ reproduce the 
virtual photon contributions to the Sirlin function~\cite{Sirlin:1967zza} 
(cf. $\Delta_{H,V}-2\pi^2/\beta$ in Ref.~\cite{Hill:2023acw}). 
Finally (after subtracting the product of the 
$O(Z\bar{\alpha})$ term of $\tilde F$ and the $O(\bar{\alpha})$ 
correction $\hat{g}+11/4$), the quantities ${\cal A}_{H,\rm 1bg}^{(2)}$ and ${\cal B}_{H,\rm 1bg}^{(2)}$
give the $Z\bar\alpha^2$ coefficient:
\begin{align}
K(w,L) &=   \Bigl( \frac{w^{2}}{s^{2}} - \frac{4 w^{2} + 1}{4 w s}\Bigr)\,\varphi^{2} + \Bigl( \frac{7 w^{2} - 5}{2 s^{2}} - \frac{13 w^{2} - 3}{6 w s}\Bigr)\,\varphi + \frac{w^{2} + 1}{s^{2}}\,L_s - \frac{2 \left(2 w^{2} - 1\right)}{s^{2}}\,L_w - \frac{w^{2} + 1}{s^{2}}\,\ln(1-x) \nonumber\\
  &\quad  + \Bigl( 1 + \frac{2 w}{3 s}\Bigr)\,L - \frac{4 w^{2} + 1}{w s}\,\mathrm{Li}_2(-x) + \frac{2 w}{s}\,\mathrm{Li}_2(-x^2) + \frac{275 w^{3} + 460 w^{2} + 159 w - 24}{36 w(w+1)^{2}} \nonumber\\
  &\quad  - \frac{2 \pi^{2} w^{2} + 48 w^{2} - 24 + \pi^{2}}{12 w s}~,
\end{align}
which agrees with Ref.~\cite{Crosas:2025xyv}. 
For $w\to\infty$ the $\varphi^2$ terms cancel, all logarithms combine into 
$L_E = \ln(2E/\mu)$
and
\begin{equation*}
K\to-\frac53\,
L_E
+\frac{131}{36}-\zeta_2=-\Bigl(\frac{2n_e}{3}+1\Bigr)L_E+\frac{8n_e}{9}+\frac{11}{4}-\zeta_2\Big|_{n_e=1},
\end{equation*}
reproducing Eq.~(18) of Ref.~\cite{Cao:2025zxs}.

\section{The $\mathcal{O}(\alpha^2)$ virtual correction to neutron decay rate}
\label{sec:HV} 

We have computed the complete two-loop virtual-virtual correction to the neutron decay rate.  These results are typically expressed in terms of on-shell $\alpha$ rather than $\bar{\alpha}$ and we follow this convention for ease of comparison with past literature. 
Let us express results in terms of the hard virtual contribution to 
the squared matrix element, as in Ref.~\cite{VanderGriend:2025mdc}: 
\begin{align} \label{eq:HV}
H_V=1+\frac{\alpha}{2\pi}H_V^{(1)}+\biggl(\frac{\alpha}{2\pi}\biggr)^2H_V^{(2)} \,,
\end{align}
where (as already emphasized) $\alpha$ is the on-shell coupling, and we take 
$n_e=1$, $\mu_{UV}=\mu$, $L=\ln(\mu/m)$. 
With $w=E/m>1$, we introduce
\begin{equation*}
s=\sqrt{w^2-1},\quad \beta=\frac{s}{w},\quad x=w-s,\quad \varphi=\ln(w+s),\quad
\ell_\pm=\ln(1\pm x^2),\quad L_w=\ln(2w),\quad L_E=L_w-L .
\end{equation*}
Note that $\varphi =\tfrac12\ln\frac{1+\beta}{1-\beta}$ when expressed in terms of the electron's 4-velocity $\beta$. 

At one-loop order,
\begin{align} \label{eq:HV1}
H_V^{(1)}&=-L+\frac{4w}{s}\varphi L+\frac{2s}{w}\varphi+\frac{2\pi^2 w}{s}-\frac{2w}{s}\Bigl[\mathrm{Li}_2(1-x^2)+\varphi^2\Bigr] \,.
\end{align}
At two loops we have
\begin{align} \label{eq:H2V}
H_V^{(2)}&=\tfrac12\bigl(H_V^{(1)}\bigr)^2+\tfrac12\,\mathcal R+\frac{\varphi^2+\pi^2}{s^2}-\frac{2\varphi^2}{w^2s^2}+\frac43\,L\,H_V^{(1)} \,.
\end{align}
where $\mathcal R\equiv \mathrm{Re}\bigl[R_{\cal A}+R_{\cal B}/w^2\bigr]$
and $R_{\cal A}$, $R_{\cal B}$ are the non-exponentiating parts of 
the two-loop coefficients: 
${\cal A}_{H}^{(2)}=\frac12({\cal A}_H^{(1)})^2+R_{\cal A}$, ${\cal B}_H^{(2)}={\cal A}_H^{(1)}{\cal B}_H^{(1)}+R_{\cal B}$.
\begin{align}
\mathrm{Re}\,R_{\mathcal A}&=\pi^2\,\frac{333w^3-153w^2-277w+113}{18(w+1)(w-1)^2}+\frac{1}{s^2}\bar a_0+\frac{w}{s}\bar a_1+\frac{w^2}{s^2}\bar a_2+\frac{w}{s^3}\bar a_3+\frac{1}{s^4}\bar a_4,\\
\mathrm{Re}\,R_{\mathcal B}&=-\pi^2\,\frac{2w(9w^2-3w-4)}{3(w+1)(w-1)^2}+\frac{w}{s}\bar b_1+\frac{w^2}{s^2}\bar b_2+\frac{w}{s^3}\bar b_3+\frac{w^2}{s^4}\bar b_4,
\end{align}
where $a_i$, $b_i$ are the functions of Eq.~(\ref{eq:AB2vals}), and
\begin{equation*}
\bar a_i(\varphi)\equiv\mathrm{Re}\,a_i(\varphi-i\pi),\qquad \bar b_i(\varphi)\equiv\mathrm{Re}\,b_i(\varphi-i\pi),\qquad
\mathrm{Li}_2(x)\to\mathrm{Li}_2(-x),\quad \ln(1+x)\to\ln(1-x),
\end{equation*}
with $\ell_\pm$ and $\mathrm{Li}_n(\pm x^2)$ unchanged. Explicitly, $\varphi^2\to\varphi^2-\pi^2$, $\varphi^3\to\varphi^3-3\pi^2\varphi$,
$\varphi^4\to\varphi^4-6\pi^2\varphi^2+\pi^4$.

At leading power in $m/E$, 
\begin{align}
[H_V^{(1)}]_0&=-2L_w^2+(4L+2)\,L_w-L+\frac{5\pi^2}{3}, \\
[H_V^{(2)}]_0&=\tfrac12[H_V^{(1)}]_0^2+\tfrac12\,U+\frac{8\pi^2}{9}(3L_E-4)+\frac43L\,[H_V^{(1)}]_0,
\end{align}
with
\begin{align}
U&=-\frac{16}{9}L_w^3+\Bigl(\frac{16L}{3}+\frac{64}{9}\Bigr)L_w^2+\Bigl(-\frac{32L^2}{3}-\frac{16L}{3}-\frac{58\pi^2}{9}-\frac{308}{27}+24\zeta_3\Bigr)L_w
\nonumber\\&\quad\quad\quad\quad\quad+\frac{20L^2}{3}+\Bigl(\frac{32\pi^2}{9}-\frac{35}{6}\Bigr)L+\frac{11663}{432}+\frac{79\pi^2}{6}-4\pi^2\ln2-18\zeta_3-\frac{9\pi^4}{10}. \label{eq:U}
\end{align}
Our results reproduce the leading corrections in the small $m/E$ limit from Ref.~\cite{VanderGriend:2025mdc}.  In particular, 
the spacelike form factor ${\cal A}_H(w,\mu)$ reduces to the $\order(\bar{\alpha}^2)$ expansion of $F_R(-w,\mu)F_J(m,\mu)F_H(-E,\mu)$ from Eq.~(16) of Ref.~\cite{VanderGriend:2025mdc}, 
and the timelike continuation of $|{\cal A}_H|^2$ reduces to 
the product of the spacelike $|{\cal A}_H|^2$ and the enhancement factor $|F_H(E,\mu_*)/F_H(E,-\mu_*-i0)|^2$ from Eq.~(14) of Ref.~\cite{VanderGriend:2025mdc}.

\section{Discussion} \label{sec:discussion}

We have computed the virtual $\alpha^2$ corrections to the neutron decay rate, with the results given by Eq.~(\ref{eq:H2V}). 
The large $E/m$ limit reproduces the results of Ref.~\cite{VanderGriend:2025mdc}.  
We have derived this result in two ways: by direct evaluation of the time-like diagrams; and by the subtracted dispersion integral (\ref{eq:dispersive}).  The latter method takes as input the zero-recoil form factor which may be identified with a matching coefficient in two-step HQET matching~\cite{Neubert:1992tg,Grozin:2005tr}; we give the first direct computation of this object, at two loops, for QED. 
The dispersive integral also takes as input the imaginary part of the background field diagrams, whose real part enters the $Z\alpha^2$ nuclear corrections (as well as the two-background field, $Z^2\alpha^2$ diagram).   

For completeness, we have computed also the real-virtual and double-real
corrections at order $\alpha^2$ for neutron beta decay (see \cref{app:real} for a discussion).
In \cref{tab:table1_fullF}, 
we compute the static-limit radiative correction, $\delta_{R,\, {\rm static}}$, at different orders in perturbation theory, and including successively smaller corrections (see Eqs.~(\ref{eq:Runres}), (\ref{eq:RF}) in Appendix~\ref{app:spectrum} for the definition of 
$\delta_{R,\, {\rm static}}$ in terms of soft and hard functions). 
In the ``without Fermi function'' column the results are expressed 
with hard function expanded as 
\begin{equation}
    H
    = 1 + \qty(\frac{\alpha}{2\pi}) H^{(1)} + \qty(\frac{\alpha}{2\pi}) ^2 H^{(2)} \,,
\end{equation}
and $H^{(2)}$ is decomposed in terms of doubly-virtual $H^{(2)}_V$, as computed above, real-virtual $H^{(2)}_{RV}$, and doubly-real corrections $H^{(2)}_{RR}$. In the ``with Fermi function $\tilde{F}$'' column we instead write 
\begin{equation}
    H
    = \tilde{F} \qty[1 + \qty(\frac{\alpha}{2\pi}) \tilde{H}^{(1)} + \qty(\frac{\alpha}{2\pi}) ^2 \tilde{H}^{(2)} ] \,,
\end{equation}
where $\tilde{F}$ is the gauge invariant 
$\overline{\rm MS}$ expression containing all powers of $Z^n \alpha^n$, cf.~\cref{FF-def} replacing $\bar{\alpha}\rightarrow \alpha$; this expression resums the leading $\pi\alpha/\beta$ correction while ensuring that there is no non-integrable singularity at $\beta\to 0$ (see \cref{app:resummation} for a discussion).  
\begin{table}[t]
\centering\small\setlength{\tabcolsep}{4pt}
\begin{tabular}{r@{\hspace{3em}}lcc}
\toprule
 & Approximation & Without Fermi function & With Fermi function $\tilde F$\\
\midrule
$\phantom{1+}H^{(1)}$
 & exact & $28.8038\pm0.08112\pm0.04858$ & $29.2301\pm0.14965\pm0.01363$\\[1pt]
\hline\\[-10pt]
\multirow{2}{*}{$\phantom{1}+H_V^{(2)}$} &  $[H_V^{(2)}]_0$ & $29.0408\pm0.04932\pm0.04908$ & $29.1972\pm0.02074\pm0.01369$\\
& exact & $29.2939\pm0.01665\pm0.04931$ & $29.1850\pm0.01785\pm0.01371$\\[1pt]
\hline\\[-10pt]
\multirow{3}{*}{$\phantom{1}+H_{RV}^{(2)}$} &  $\frac{\pi\alpha}{\beta} H_R^{(1)}$ & $29.1863\pm0.01714\pm0.01283$ & $29.1850\pm0.01785\pm0.01371$\\
& $H_V^{(1)}H_R^{(1)}+\frac12\bigl(H_R^{(1)}\bigr)^2$
& $29.2021\pm0.00081\pm0.00048$ & $29.2012\pm0.00060\pm0.00002$\\
& exact (numerical) & $29.1995\pm0.00080\pm0.00048$ & $29.1986\pm0.00059\pm0.00003$\\[1pt]
\hline\\[-10pt]
$\phantom{1}+H_{RR}^{(2)}$ & exact (numerical) & $29.1992\pm0.00079\pm0.00047$ & $29.1983\pm0.00059\pm0.00003$\\
\bottomrule
\end{tabular}
\caption{Contributions to the static-limit radiative correction to the neutron lifetime, $\delta_{R,\rm static}(\mu_{\rm UV}=\Delta)$, in units of $10^{-3}$.  First uncertainty: variation of $\mu\in[m/2,\,2\Delta]$; second uncertainty: variation of the soft-photon parameter 
$\Lambda_\gamma
\in[\Delta/2,\,2\Delta]$ (see \cref{app:real}). Right column: $H=\tilde F\,\tilde H$ with $\tilde F$ to all orders. Both results are consistent with $\delta_{R,{\rm static}}= (29.199\pm 0.001)\times 10^{-3}$, which we take as our new value for neutron beta decay. Notice that the RV and RR corrections not captured by the $(\pi \alpha/\beta)\times H_R^{(1)}$ term only enter at $\sim 2\times10^{-5}$ such that the virtual corrections (including the proper electron-mass dependence) and proper factorization of  $(1+ \pi \alpha/\beta)$ are the major drivers of the $\mathcal{O}(\alpha^2)$ correction.
\label{tab:table1_fullF}}
\end{table}

We further show $H_V^{(2)}$ first with its $m\to 0$ limit, then with its full mass dependence.  We show 
$H_{RV}^{(2)}$ just with its Coulomb-enhanced contribution (automatically included already with $H^{(1)}$ in the right-hand column); with all exponentiating terms from first-order corrections; and with the full correction.  
Referring to the numerical results as displayed in \cref{tab:table1_fullF}, we thus update our old result 
$\delta_{R,\,\rm static}(\mu_{\rm UV}=\Delta)= (29.18\pm 0.07 \pm 0.01 \pm 0.02) \times 10^{-3}$ \cite{VanderGriend:2025mdc} to:
\begin{align}
    \delta_{R,\,\rm static}(\mu_{\rm UV}=\Delta)= ( 29.199 \pm 0.001 )\times 10^{-3} \,, 
\end{align}
eliminating the largest source of theoretical uncertainty in this quantity, and 
pushing the error well below experimental precision.\!\footnote{Uncertainties on $g_A/g_V$~\cite{Markisch:2018ndu} and the short-distance corrections~\cite{Seng:2018yzq,Czarnecki:2019mwq,Ma:2023kfr,Cirigliano:2023fnz} are still the dominant sources of uncertainty.  For definiteness, we have taken the first column of Table~\ref{tab:table1_fullF} as central value.  The uncertainty $\pm 0.001 \times 10^{-3}$ encompasses both the uncertainty of this column one value, and the difference between column one and column two.}

\begin{table}[t]\centering\small
\begin{tabular}{lccc|c|cccc|c}\toprule
Parent & $\beta^\pm$ & $W_0$ & $Z$ & $\left\langle \delta^{(1,0)}\right\rangle $ & VV & RV & RR & $e^+e^-$ & $\left\langle \delta^{(2,0)}\right\rangle $ \\\midrule
$n$ & $\beta^-$ & 2.5310 & -1 & -2.2846 & +0.0184 & -0.0158 & +0.0106 & --- & +0.0132\\
${}^3$H & $\beta^-$ & 1.0364 & -2 & -2.2936 & +0.0058 & +0.0004 & +0.0000 & --- & +0.0062\\
${}^{10}$C & $\beta^+$ & 2.7339 & +5 & -2.3933 & +0.0209 & -0.0207 & +0.0131 & --- & +0.0133\\
${}^{14}$O & $\beta^+$ & 4.5406 & +7 & -2.4402 & +0.0507 & -0.0703 & +0.0331 & +0.0000 & +0.0134\\
${}^{22}$Mg & $\beta^+$ & 7.0715 & +11 & -2.5324 & +0.1049 & -0.1522 & +0.0607 & +0.0000 & +0.0135\\
${}^{26m}$Al & $\beta^+$ & 7.2831 & +12 & -2.5444 & +0.1099 & -0.1593 & +0.0629 & +0.0000 & +0.0135\\
${}^{26}$Si & $\beta^+$ & 8.4733 & +13 & -2.5650 & +0.1383 & -0.1986 & +0.0747 & +0.0001 & +0.0144\\
${}^{34}$Cl & $\beta^+$ & 9.7469 & +16 & -2.5908 & +0.1697 & -0.2405 & +0.0866 & +0.0002 & +0.0159\\
${}^{34}$Ar & $\beta^+$ & 10.8628 & +17 & -2.5984 & +0.1976 & -0.2766 & +0.0963 & +0.0003 & +0.0177\\
${}^{38m}$K & $\beta^+$ & 10.8283 & +18 & -2.6008 & +0.1968 & -0.2755 & +0.0960 & +0.0003 & +0.0176\\
${}^{38}$Ca & $\beta^+$ & 11.9396 & +19 & -2.6033 & +0.2249 & -0.3107 & +0.1053 & +0.0004 & +0.0199\\
${}^{42}$Sc & $\beta^+$ & 11.5759 & +20 & -2.6040 & +0.2157 & -0.2993 & +0.1023 & +0.0003 & +0.0191\\
${}^{46}$V & $\beta^+$ & 12.8013 & +22 & -2.6006 & +0.2469 & -0.3374 & +0.1120 & +0.0005 & +0.0219\\
${}^{50}$Mn & $\beta^+$ & 13.9402 & +24 & -2.5905 & +0.2759 & -0.3721 & +0.1204 & +0.0006 & +0.0249\\
${}^{54}$Co & $\beta^+$ & 15.1338 & +26 & -2.5737 & +0.3063 & -0.4074 & +0.1287 & +0.0008 & +0.0285\\
${}^{62}$Ga & $\beta^+$ & 16.9669 & +30 & -2.5257 & +0.3528 & -0.4597 & +0.1404 & +0.0011 & +0.0347\\
${}^{74}$Rb & $\beta^+$ & 19.3852 & +36 & -2.4154 & +0.4133 & -0.5246 & +0.1539 & +0.0016 & +0.0442\\
\bottomrule\end{tabular}
\caption{$Z^0$ contributions of order $\alpha$ and $\alpha^2$ to nuclear $\beta$ decay, in units of $10^{-3}$, averaged with the Fermi function [$\tilde F(Z\alpha,E,\mu_L)$, Eq.~(22) of Ref.~\cite{Cao:2025zxs} with $\xi=-Z\alpha E/p$, $Z=+Z_{\rm daughter}$ ($\beta^+$) and $Z=-Z_{\rm daughter}$ ($\beta^-$) ].
The second order correction is broken up into virtual-virtual, real-virtual, and real-real whose sum gives the total $\mathcal{O}(\alpha^2)$ correction.  The maximum electron/positron energies in units of $m_e$, $W_0$, are taken from Ref.~\cite{Hardy:2014qxa} (see Ref.~\cite{Hardy:2020qwl} for the 2020 update of the survey).  Following Ref.~\cite{Cao:2025zxs} we take $\mu_L=2 E_m$, with $E_m=m_e W_0$. 
\label{tab:nuc} }
\end{table}

Our results may be applied also to superallowed nuclear beta decays of interest to $|V_{ud}|$ determinations by writing (again using on-shell $\alpha$ to match literature conventions), 
\begin{equation}
    \begin{split}
    \label{factorized-form-Gamma-complete-onshell}
    \!\!\!\! \Gamma = |\mathcal{C}(\mu_L)|^2 \!\int \dd \Phi_e \tilde{F}(Z \alpha ,E,\mu_L)  \bigg[1 + \delta^{(1,0)}(E,\mu_L) &+  \delta^{(2,1)}(E,\mu_L)  + \delta^{(2,0)}(E,\mu_L) + \delta^{(3,2)}(E,\mu_L) + \ldots  \bigg]~,
    \end{split}
\end{equation}
where $\dd \Phi_e = \dd E_e ~p_e E_e (W_0-E_e)^2$ is the tree-level phase space, $\delta^{(m,n)}$ is the $Z^n\alpha^m$ correction at fixed electron kinematics (including integration over the real radiation's phase space), $\mathcal{C}(\mu_L)$ is the low-energy vector-current Wilson coefficient which contains all effects related to electroweak dynamics, hadronic, and nuclear structure (including large logarithms).  One can then write the integrated decay rate as 
\begin{equation}
     \Gamma = |\mathcal{C}(\mu_L)|^2 \qty[\int \dd \Phi_e ~\tilde{F}(Z \alpha ,E,\mu_L)] \times \qty[ 1+ \langle\delta^{(1,0)} \rangle(\mu_L) +   \langle\delta^{(2,1)}\rangle(\mu_L)  +  \langle\delta^{(2,0)} \rangle(\mu_L) + \ldots  ]~,
\end{equation}
and $|\mathcal{C}(\mu_L)|^2$ can be written in terms of $|\mathcal{C}(\mu_H)|^2$ multiplying an RG-kernel \cite{Hill:2023acw,Borah:2024ghn,Cao:2025zxs}. Using the results of the present paper we can determine $\delta^{(2,0)}(E,\mu_L)$ averaged over phase space (including the Fermi function) which then enters as part of the radiative corrections to superallowed beta decay lifetimes. We quote our value for a scale choice of $\mu_L=2m_eW_0$.

Table~\ref{tab:nuc} displays representative nuclei undergoing $0^+ \to 0^+$ transitions ($\beta^+$ emitters), as well as the neutron and $^3{\rm H}$ cases for comparison ($\beta^-$ emitters).  The columns show the maximum electron/positron energy $W_0$, the $\order(\alpha)$ correction for comparison, and the purely virtual, real-virtual and purely real contributions to the second order correction.  
In the standard analysis of superallowed decays~\cite{Hardy:2020qwl}, the Z-independent $O(\alpha^2)$
terms enter only through leading logarithms (cf.\ Eq.~(55) of Ref.~\cite{Cao:2025zxs}); the values of
$\langle\delta^{(2,0)}\rangle$ in Table~\ref{tab:nuc}, evaluated at $\mu_L=2E_m$, are therefore an
additional contribution to $\delta_R'$.
For the superallowed transitions, the contributions are larger than $10^{-4}$, with numerical cancellations resulting in a total $\sim (0.1-0.5) \times 10^{-4}$.  This is comparable to experimental uncertainty in $|V_{ud}|$ extractions~\cite{Hardy:2020qwl}. These $Z$-independent terms complement recent effective field theory analyses of the nuclear-structure-dependent corrections~\cite{Cirigliano:2024rfk,Cirigliano:2024nfi,Novario:2026vud,Novario:2026dns}.
For the neutron, the entries of Table~\ref{tab:nuc} combine with $\langle\tilde F\rangle-1=34.160\times10^{-3}$ and
$\langle\delta^{(2,1)}\rangle=-0.105\times10^{-3}$ to $\delta_{R,\rm static}=31.702\times10^{-3}$ at $\mu_{UV}=2W_0$, i.e.\
$29.198\times10^{-3}$ at $\mu_{UV}=\Delta$, in agreement with Table~\ref{tab:table1_fullF}.

In summary: We have derived two-loop results for nuclear and nucleon beta decay.  
Our results provide complete $\order(\alpha^2)$ corrections for rate and electron energy spectra. 
The $Z\alpha^2$ correction to the nuclear beta decay spectrum involves the real parts of the background field diagrams in Eq.~(\ref{eq:Mdiagrams}).  
Our result $K(w,L)$ reproduces the large $E/m$ limit from Ref.~\cite{Cao:2025zxs}.  The complete electron mass dependence is in agreement with the recent result of Crosas and Mereghetti~\cite{Crosas:2025xyv}.%
\footnote{Cf. Equation (3.3) of Ref.~\cite{Crosas:2025xyv}, taking the plus sign for the positron convention used here.} The $\alpha^2$ results derived herein are new, and we have checked their consistency with the $Z\alpha^2$ results via a dispersion relation. Finally, we note that the $\alpha^2$ results may also be applied to the nuclear case, subleading in $Z$ compared to the $Z^2$ and $Z^1$ corrections at two-loop order, but potentially of interest for precision extractions of $|V_{ud}|$. In the power counting of \cite{Cao:2025zxs}, this leaves only the finite $Z^2\alpha^3$ low-energy matrix element as a missing ingredient for a complete $O(\alpha^2)$ analysis (the RG resummation is known at this order).  

\vskip 0.1in
\noindent{\bf Acknowledgements}
\vskip 0.1in
\noindent
This work was supported by the U.S. Department of Energy, Office of Science, Office of High Energy Physics, under Award DE-SC0019095.
PVG was supported by the U.S. Department of Energy under Contract No. DE-SC0012704. This work was produced by Fermi Forward Discovery Group, LLC under Contract No. 89243024CSC000002 with the U.S. Department of Energy, Office of Science, Office of High Energy Physics.
The AI model Claude (Anthropic) was used to assist with code development, numerical checks and the drafting and editing of parts of the appendices; see the paragraph on the use of AI tools in Appendix~\ref{app:integrals}.
\newpage

\appendix
\crefalias{section}{appendix}

\section{Evaluation of the two-loop integrals}
\label{app:integrals}

We describe here the evaluation of two-loop integrals, including the background field diagrams of Eq.~(\ref{eq:Mdiagrams}), and the heavy-light form factor diagrams of Eq.~(\ref{eq:Mwdiagrams}).  

\subsection{Projection onto form factors}
After loop integration the Dirac string between
$\bar u(p)$ and $\Gamma h_v$ has the form $N=a_1+a_2\slashed v+a_3\slashed p+a_4\slashed v\slashed p$. The $a_i$ are obtained
\emph{before} integration by traces against $\{1,\slashed v,\slashed p,\slashed v\slashed p\}$. Using $\bar u(p)\slashed p=m\,\bar u(p)$ gives
\begin{equation}
\bar u(p)\,N\,\Gamma h_v=\bar u(p)\,\Bigl[\mathcal A+\frac{\slashed v}{w}\,\mathcal B\Bigr]\,\Gamma h_v,\qquad
\mathcal A=a_1+m a_3+2mw\,a_4,\qquad \mathcal B=w\,(a_2-m a_4) .
\end{equation}
The Gram matrix of the basis is singular only at $w=1$, where $\mathcal A+\mathcal B$ reduces to $\tfrac14\mathrm{Tr}[(1+\slashed v)N]$. 

\subsection{Integral families and IBP reduction}
We set $m=1$, $v^2=1$, $p^2=1$, $v\cdot p=w$ and $d=4-2\epsilon$, and use loop momenta $k,q$. Master integrals are quoted with the measure
$\int d^dl/(i\pi^{d/2})$ per loop and no further factor. $\overline{\rm MS}$ quantities are defined with a factor $\Gamma(1+\epsilon)$ per loop, as in the main text;
the renormalized form factors $\mathcal A_H$ and $\mathcal B_H$ are the same if $e^{-\gamma_E\epsilon}$ is used instead.
The denominators follow Lee's convention (all $-i0$). The nine denominators
\begin{align}
&D_1=-2v\!\cdot\!k,\quad D_2=-2v\!\cdot\!q,\quad D_3=-2v\!\cdot\!(k+q),\quad D_4=-k^2,\quad D_5=-q^2,\quad D_6=-(k+q)^2,\nonumber\\
&D_7=1-(k+p)^2,\quad D_8=1-(q+p)^2,\quad D_9=1-(k+q+p)^2 , 
\end{align}
obey two linear relations, $D_3=D_1+D_2$ and $D_9=D_6+D_7+D_8-D_4-D_5$. LiteRed2~\cite{Lee:2012cn,Lee:2013mka} (\texttt{NewDsSet} with partial
fractioning) splits them into ten seven-denominator families \texttt{hlA01}--\texttt{hlA10}. The three that contain all
masters are
\begin{align}
\texttt{hlA01}&:\ \{-2v\!\cdot\!k,\,-2v\!\cdot\!(k+q),\,-k^2,\,-q^2,\,-(k+q)^2,\,1-(k+p)^2,\,1-(k+q+p)^2\},\nonumber\\
\texttt{hlA02}&:\ \{-2v\!\cdot\!k,\,-2v\!\cdot\!(k+q),\,-k^2,\,-q^2,\,-(k+q)^2,\,1-(q+p)^2,\,1-(k+q+p)^2\},\\
\texttt{hlA03}&:\ \{-2v\!\cdot\!k,\,-2v\!\cdot\!(k+q),\,-k^2,\,-q^2,\,1-(k+p)^2,\,1-(q+p)^2,\,1-(k+q+p)^2\}.\nonumber
\end{align}
A further family \texttt{hlF01} covers the massive vacuum-polarization loop, and \texttt{hlOne01} $=\{-2v\!\cdot\!k,-k^2,1-(k+p)^2\}$ the one-loop case.
IBP reduction (\texttt{GenerateIBP}, \texttt{FindSymmetries}, \texttt{FindExtSymmetries}, \texttt{SolvejSector}) expresses all diagrams through
18 two-loop masters $J=j[\mathrm{family},n_1,\dots,n_7]$:
\begin{center}\small
\begin{tabular}{ll}
\texttt{hlA01}: & $(0,0,0,0,0,1,1)$, $(0,0,0,1,1,1,0)$, $(0,1,0,0,0,1,1)$, $(0,1,0,1,0,1,0)$, $(0,2,0,1,0,1,0)$, $(0,1,0,1,1,1,0)$,\\
 & $(0,1,1,1,0,0,1)$, $(1,1,0,0,0,1,1)$, $(0,1,1,1,0,1,1)$, $(1,1,0,1,1,1,0)$, $(0,2,1,1,0,1,1)$\\
\texttt{hlA02}: & $(1,0,0,0,0,1,1)$, $(1,0,0,1,0,1,1)$, $(1,1,0,0,0,1,1)$, $(2,0,0,1,0,1,1)$\\
\texttt{hlA03}: & $(0,0,0,0,1,1,1)$, $(1,0,0,0,1,1,1)$, $(2,0,0,0,1,1,1)$
\end{tabular}\end{center}
The one-loop masters are $j[\texttt{hlOne01},0,0,1]=\Gamma(\epsilon-1)$ and
$j[\texttt{hlOne01},1,0,1]=\Gamma(\epsilon)\,x\,{}_2F_1(\epsilon,1;2\epsilon;1-x^2)/(2\epsilon-1)$, where $x=w-\sqrt{w^2-1}$.

\subsection{Differential equation and canonical form}
Differentiating with respect to $w$ (LiteRed \texttt{Dinv}) and reducing gives a closed system $\partial_wJ=M_w(\epsilon,w)J$ for the 18 masters.
The denominator factors of $M_w$ are $w$ and $w\pm1$. The variable
\begin{equation}
x=w-\sqrt{w^2-1}\in(0,1)\ \ (w>1),\qquad w=\tfrac12(x+x^{-1}),
\end{equation}
rationalizes the square root. A transformation $J=T(\epsilon,x)\,g$ brings the system to canonical form, which we verify exactly
($M_xT-\partial_xT=T M_\epsilon$) and independently with Libra~\cite{Lee:2014ioa,Lee:2020zfb} (\texttt{EFormQ}):
\begin{equation}
\frac{dg}{dx}=\epsilon\Bigl(\frac{A_0}{x}+\frac{A_1}{x-1}+\frac{A_2}{x+1}+\frac{2x\,A_3}{1+x^2}\Bigr)g .
\end{equation}
The alphabet is therefore $\{x,\,1-x,\,1+x,\,1+x^2\}$; the letter $1+x^2=2wx$ produces the arguments $\pm x^2$ in Eqs.~(\ref{eq:AB2}),(\ref{eq:AB2vals}). The matrices $A_i$
are rational numbers and are given in the supplementary file \texttt{epsform\_w.m}.

\subsection{Boundary conditions and solution}
The boundary is taken at $x=1$ ($w=1$, zero recoil). There $p=m\,v$, and the masters become single-scale integrals, pure functions of $\epsilon$.
Five are needed. Four are products of Gamma functions:
\begin{align}
&\Gamma(\epsilon-1)^2 &&\text{two massive tadpoles},\nonumber\\
&\frac{\sqrt\pi}{2}\frac{\Gamma(2\epsilon-\frac32)\Gamma(\epsilon-\frac12)^2}{\Gamma(2\epsilon-1)} &&\text{static line $v\!\cdot\!k$, massive lines $q$ and $k+q$},\nonumber\\
&\frac{\Gamma(\epsilon)\Gamma(1-\epsilon)^2\Gamma(2\epsilon-1)\Gamma(3-4\epsilon)}{\Gamma(2-2\epsilon)\Gamma(3-3\epsilon)} &&\text{sunset, one massive and two massless lines, $p^2=m^2$},\label{eq:bcGamma}\\
&\frac{\Gamma(2\epsilon-1)\Gamma(1-\epsilon)\Gamma(3\epsilon-2)\Gamma(4\epsilon-3)}{\Gamma(6\epsilon-4)} &&\text{static line $v\!\cdot\!k$, massless line $q$, massive line $k+q$}.\nonumber
\end{align}
The fifth is the sunset with three massive lines at $p^2=m^2$,
\begin{align}
B(\epsilon)&=2^{1-2\epsilon}\sqrt\pi\,\biggl[\frac{\Gamma(\epsilon)\Gamma(1-\epsilon)\Gamma(2\epsilon-1)}{\Gamma(\epsilon+\frac12)\Gamma(2-\epsilon)}\,
{}_3F_2\bigl(2\epsilon-1,\tfrac12,1;\epsilon+\tfrac12,2-\epsilon;1\bigr)\nonumber\\
&\qquad\qquad+\frac{4^{\epsilon-1}\Gamma(\epsilon-1)\Gamma(\epsilon)}{\Gamma(\frac32)}\,{}_3F_2\bigl(\epsilon,\tfrac32-\epsilon,1;\tfrac32,3-2\epsilon;1\bigr)\biggr],\\
e^{2\epsilon\gamma_E}B(\epsilon)&=-\frac3{2\epsilon^2}-\frac{17}{4\epsilon}-\frac{59}8-\frac{\pi^2}4+\Bigl(-\frac{65}{16}-\frac{49\pi^2}{24}+\zeta_3\Bigr)\epsilon\nonumber\\
&\quad+\Bigl(\frac{1117}{32}-\frac{475\pi^2}{48}-\frac{151}6\zeta_3+8\pi^2\ln2-\frac{7\pi^4}{240}\Bigr)\epsilon^2+O(\epsilon^3).\label{eq:bcSunset}
\end{align}
The coefficients of this expansion were identified by PSLQ~\cite{Ferguson:1999pslq} from high-precision evaluations of the ${}_3F_2$ form, and agree with it to at least 20 digits.
Terms beyond $\epsilon^2$ do not contribute to the finite form factors. All $w=1$ masters were also checked with
pySecDec~\cite{Borowka:2017idc}. The canonical system is integrated order by order in $\epsilon$ up to weight four, with the
boundary at $x=1$. The result is expressed through Goncharov polylogarithms $G(a_1,\dots,a_n;y)$ with $y=1-x$ and letters $\{0,1,2,1\pm i\}$,
and then converted to the classical polylogarithms $\mathrm{Li}_n(x)$, $\mathrm{Li}_n(\pm x^2)$ of Eqs.~(\ref{eq:AB2}),(\ref{eq:AB2vals}).

Checks of the bare form factors:
\begin{itemize}
\item The spurious $\epsilon^{-4}$ and $\epsilon^{-3}$ poles cancel identically, for every charge structure.
\item In the limit $w\to 1$, $\mathcal A+\mathcal B$ reproduces
a separate computation performed directly at $w=1$.
That computation has its own integral families and reduction, and shares only the boundary constants of Eqs.~(\ref{eq:bcGamma})--(\ref{eq:bcSunset}).
\end{itemize}

\subsection{Renormalization}

The renormalized coefficients are $\mathcal M(\mu)=Z_SZ_H^{-1}Z_\psi^{1/2}Z_h^{1/2}\mathcal M_{\rm bare}$, cf. Eq.~(\ref{eq:MmuMbare}). The factors are:
\begin{itemize}
\item $Z_\psi^{1/2}$ and $Z_h^{1/2}$: the on-shell wave-function renormalization of the electron and of the static line, together with the on-shell mass counterterm;
\item $Z_H^{-1}$: the $\overline{\rm MS}$ renormalization of the heavy-light current; 
\item $Z_S$: the operator renormalization constant of the soft theory, i.e.\ of two static lines with relative velocity $w$. It is defined in $\overline{\rm MS}$ with the
zero-flavor coupling $\alpha'$, and removes the remaining infrared (cusp) poles:
\begin{equation}
Z_S(w)=\exp\Bigl[-\frac{2a'}{\epsilon}\bigl(w\,r(w)-1\bigr)\Bigr],\qquad a'=\frac{\alpha'}{4\pi},\qquad w\,r(w)=\frac{w\varphi}{s},
\end{equation}
with $\alpha'=\bar\alpha[1-\frac{\bar\alpha}{3\pi}(e^{\epsilon\gamma_E}\Gamma(1+\epsilon)(\mu^2/m^2)^\epsilon-1)/\epsilon]$.
\end{itemize}
In addition, the bare coupling is expressed through $\bar\alpha(\mu)$ with $n_e=1$ dynamical electron.
All poles cancel. This holds at $L=0$ and $L\neq0$.
At $w=1$ the result reproduces $E(m)$ of Eq.~(\ref{eq:Em}). 

\subsection{Background-field (cut) integrals}

For the dispersive computation (Sec.~\ref{sec:dispersive}) the one-background-photon diagrams are evaluated in the family
\begin{equation}
\texttt{famC}:\ \{\,v\!\cdot\!q\ (\hbox{cut}),\ k^2,\ q^2,\ -v\!\cdot\!k,\ (k+q-p)^2-m^2,\ (q-p)^2-m^2,\ (k-p)^2-m^2\,\},
\end{equation}
where the cut denotes $2\pi\delta(v\!\cdot\!q)$, Eq.~(\ref{eq:deltacut}). The cut is declared to LiteRed2 (\texttt{CutDs}), so that IBP relations and symmetries respect it. 
With it the reduction yields 14 cut masters. Three masters contain a squared propagator. For two of them we use an IBP identity to trade LiteRed's default master, with $k^2$ squared,
for the integral with the fifth denominator, $(k+q-p)^2-m^2$, squared.

The $14\times14$ differential equation in $w$ is written in $y$, with $w=\frac12(y+y^{-1})$. Libra then brings it to $\epsilon$-form in three steps:
\begin{itemize}
\item normalization of the $1\times1$ blocks and explicit transformations for the three $2\times2$ blocks;
\item \texttt{Fuchsify} for the off-diagonal part;
\item \texttt{FactorOut}.
\end{itemize}
The solution is fixed by its behaviour at $w\to1$. With $x=y-1$, \texttt{GetLcs} shows that four expansion coefficients are needed:
\begin{itemize}
\item $c_{157}$: the $x^0$ term of $j[\texttt{famC},1,0,0,0,1,0,1]$;
\item $c_1$: the $x^{1-2\epsilon}$ term of $j[\texttt{famC},1,0,0,0,0,1,1]$;
\item $c_2$: the $x^0$ term of $j[\texttt{famC},1,1,0,0,1,0,0]$;
\item $c_3$: the $x^{3-6\epsilon}$ term of $j[\texttt{famC},1,1,0,0,2,0,0]$.
\end{itemize}
In the normalization $\int\!\!\int d^dk\,d^dq/(2\pi)^{2d}(-2\pi i)\delta(v\!\cdot\!q)$, with $m=1$ and $\mathcal N=2^{4\epsilon-7}\pi^{2\epsilon-7/2}$, they are
\begin{align}
c_{157}&=-\mathcal N\,\frac{\Gamma(\epsilon-\frac12)^2\Gamma(2-2\epsilon)\tan(2\pi\epsilon)}{\Gamma(\frac52-2\epsilon)},\qquad
c_1=-i\mathcal N\,\Gamma(\epsilon-\tfrac12)\Gamma(\epsilon-1)e^{i\pi\epsilon},\nonumber\\
c_2&=\mathcal N\,\frac{\Gamma(2\epsilon-\frac32)\Gamma(\epsilon-\frac12)\Gamma(\frac52-3\epsilon)}{\Gamma(2-2\epsilon)},\qquad
c_3=\mathcal N\,\frac{2^{2\epsilon-2}}{\sqrt\pi}e^{i\pi\epsilon}\Gamma(\epsilon-\tfrac12)\Gamma(\epsilon+\tfrac12)\,K_3,\\
K_3&=e^{i\pi(\frac12+\epsilon)}\bigl[B(2-2\epsilon,\tfrac12-\epsilon)+B(2-2\epsilon,-\tfrac32+3\epsilon)\bigr]+B(-\tfrac32+3\epsilon,\tfrac12-\epsilon).\nonumber
\end{align}
The solution is constructed as a path-ordered exponential (\texttt{PexpExpansion}) through weight four and expressed in HPLs~\cite{Remiddi:1999ew}. The FORM~\cite{Vermaseren:2000nd,Ruijl:2017dtg} numerators of the
five cut diagrams are mapped onto \texttt{famC}. The infrared and Coulomb subtractions of Eq.~(\ref{eq:sub}) are added analytically. The real and imaginary
parts of the one-background-photon amplitude, Eqs.~(\ref{eq:ReA1bg})--(\ref{eq:ImB1bg}), follow, and inserting them in the dispersion relation (\ref{eq:dispersive}) reproduces the direct computation.

\paragraph{Code.} The supplementary material contains the scripts in the folders below, with a \texttt{README} that gives the order in which to run them.
They require Mathematica with FeynCalc~\cite{Mertig:1990an,Shtabovenko:2016sxi,Shtabovenko:2023idz,Shtabovenko:2025fc}, LiteRed2 and Libra, and FORM for the cut numerators.
\begin{itemize}
\item \texttt{two\_loop\_w/}, the direct computation at general $w$: \texttt{HLsetup\_w.wl}, \texttt{stageA\_w.wl}, \texttt{stageB\_w.wl}, \texttt{DE\_w.wl}, \texttt{eps\_w.wl}, \texttt{FF2L\_w.wl}, \texttt{Ren\_w.wl}.
The helper files are \texttt{HLdefs\_w.wl}, \texttt{HLreduce\_w.wl}, \texttt{FCtoLR\_w.wl}, \texttt{OneLoop\_w.wl} and \texttt{check1L\_w.wl}.
The data files are \texttt{epsT\_w.m} (transformation to canonical form), \texttt{epsform\_w.m} (the matrices $A_i$) and \texttt{solution\_w.m} (the solution in Goncharov polylogarithms).
\item \texttt{two\_loop\_static/}, zero recoil: \texttt{HLsetup.wl}, \texttt{stageA.wl}, \texttt{HLreduce.wl}, \texttt{HLmasters.wl}, with the helper files \texttt{HLdefs.wl} and \texttt{FCtoLR.wl}
and the FORM file \texttt{static\_numerators.frm}.
\item \texttt{two\_loop\_cut/}, the cut family: \texttt{build\_famC.m}, \texttt{setupC.m}, \texttt{workflowC.m}, \texttt{finiteC.m}, \texttt{FiniteTotalC.m} (subtraction terms and finite amplitude), \texttt{ImClassicalC.m}, with the helper files \texttt{cutEvalC.m} and \texttt{form2wfC.m},
the FORM file \texttt{cut\_numerators.frm} and its output \texttt{cut\_numerators.out}.
\item \texttt{results/}, the results in machine-readable form: \texttt{CompactC.m} ($\mathcal A_H$, $\mathcal B_H$, Eqs.~(\ref{eq:AB1})--(\ref{eq:AB2vals})),
\texttt{HardH.m} ($H_V$, Eqs.~(\ref{eq:HV})--(\ref{eq:U})) and \texttt{BkgSplit.m} (Eqs.~(\ref{eq:A2bg})--(\ref{eq:ImB1bg})).
\item \texttt{numerics/}: real radiation, the electron spectrum and the scripts for Tables~\ref{tab:table1_fullF} and \ref{tab:nuc} (Python), see Apps.~\ref{app:real} and \ref{app:spectrum}.
\end{itemize}

\paragraph{Use of AI tools.} Parts of the supplementary code (in particular the scripts in \texttt{numerics/} and the drivers and checks of the Mathematica chains), the evaluation of the real $e^+e^-$ pair-emission contribution including the exchange term, and several of the numerical cross-checks described in these appendices were developed with the assistance of the AI model Claude (Anthropic; Claude Opus 5.5). The authors directed this work and verified the results, by comparison with independent analytic results and numerical evaluations as described in the text, and by rerunning all scripts.

\section{Real-virtual and double-real radiation}
\label{app:real}

File names in this appendix refer to the folder \texttt{numerics/} of the supplementary material, whose \texttt{README} gives the order in which to run the scripts. The scripts are written in Python and use NumPy~\cite{Harris:2020xlr}, SciPy~\cite{Virtanen:2019joe}, SymPy~\cite{Meurer:2017yhf} and mpmath~\cite{mpmath}.

In order to obtain finite results the phase space was split using a soft-photon cutoff, $\epsilon_\gamma$, in the rest-frame of the parent $\beta$-emitter. This scheme is the same one used in Refs.~\cite{VanderGriend:2025mdc,Cao:2025lrw} to estimate higher order missing real-emission corrections; we use it in part for ease of comparison with the results of Ref.~\cite{VanderGriend:2025mdc}. Above $\epsilon_\gamma$ explicit real emissions are retained whereas below $\epsilon_\gamma$ soft-photon exponentiation is assumed. 
One finds that would-be infrared divergent quantities depend on $\ln(\Delta/\epsilon_\gamma)$; we replace
$\epsilon_\gamma\to\Lambda_\gamma$ in these logarithms and vary $\Lambda_\gamma\in[\Delta/2,2\Delta]$ (see \cite{Cao:2025lrw} for a discussion of this choice).
Although this soft-photon cutoff is evaluated for energies close to $\Delta$ (and therefore ceases to be soft) it serves as a useful diagnostic of missing real-emission corrections. 

\subsection{Setup and decomposition}
We use units $m=1$ and $a=\alpha/2\pi$, with $\alpha$ on shell. The proton is static ($v=(1,\vec0)$, charge $+1$, 
propagator $i/(v\cdot l+i0)$).
The electron has $p=(E,0,0,P)$, $P=\sqrt{E^2-1}$, $\beta=P/E$, $L_\beta=\ln\frac{1+\beta}{1-\beta}$, and the endpoint is $\Delta=(m_n-m_p)/m_e$, with $K_m=\Delta-E$.
Real photons ($k$, $k^0=k$) are summed over physical polarizations in the rest frame of the proton,
\begin{equation}
\sum_{\rm pol}\varepsilon^\mu\varepsilon^{*\nu}=-g^{\mu\nu}-\frac{k^\mu k^\nu}{(v\cdot k)^2}+\frac{k^\mu v^\nu+v^\mu k^\nu}{v\cdot k},
\end{equation}
so $v\cdot\varepsilon=0$ and emission from the static line vanishes identically, also inside loops. The lepton traces are taken with the
$V\!-\!A$ current and averaged over the neutrino direction. After this average every trace is proportional to $1+3\lambda^2$. The real-radiation
functions below are therefore the same for Fermi and Gamow--Teller transitions, and independent of $\lambda=g_A/g_V$.

The hard function is $H=1+aH^{(1)}+a^2H^{(2)}$, with $H^{(1)}=H_V^{(1)}+H_R^{(1)}$ and
\begin{equation}
H^{(2)}=\underbrace{H_V^{(2)}}_{\rm VV}+\underbrace{H_V^{(1)}H_R^{(1)}+\Delta_{RV}}_{\rm RV}+\underbrace{\tfrac12\bigl(H_R^{(1)}\bigr)^2+D+\Delta_{PS}}_{\rm RR}.
\label{eq:H2decomp}
\end{equation}
The terms $H_V^{(1)}H_R^{(1)}+\frac12(H_R^{(1)})^2$ are the real-photon part of $\frac12(H^{(1)})^2$, i.e.\ of the exponentiation of the first
order. $\Delta_{RV}$, $D$ and $\Delta_{PS}$ are the non-exponentiating remainders. They are infrared finite and independent of $\mu$, $\mu_{UV}$ and the photon-energy
cutoff $\epsilon_\gamma$. With the photons sliced at $\epsilon_\gamma$ consistently with the soft function $S$ of Ref.~\cite{VanderGriend:2025mdc}, the one-photon real function is
\begin{equation}
H_R^{(1)}=\ln\frac{\epsilon_\gamma}{K_m}\Bigl(4-\frac{2L_\beta}{\beta}\Bigr)+\frac{L_\beta}{\beta}\Bigl(\frac{K_m^2}{12E^2}+\frac{2K_m}{3E}-3\Bigr)-\frac{4K_m}{3E}+6,
\end{equation}
and the soft function, exponentiated to all orders, is
\begin{equation}
S=\exp\Bigl\{a\Bigl[\Bigl(\frac{2L_\beta}{\beta}-4\Bigr)\ln\frac{2\epsilon_\gamma}{\mu}-\frac{L_\beta^2}{2\beta}-\frac2\beta\mathrm{Li}_2\Bigl(\frac{2\beta}{1+\beta}\Bigr)+\frac{L_\beta}\beta+2\Bigr]\Bigr\}.
\end{equation}

\subsection{Real-virtual remainder $\Delta_{RV}$}
\paragraph{Definition.} Let $M_0^\gamma$ ($M_1^\gamma$) be the tree (one-loop) amplitude with one real photon, and $M_0$ ($M_1$) the amplitude without a photon. Then
\begin{equation}
\delta(E,k,\theta)=\frac{\sum_{\rm pol}2\,\mathrm{Re}\,M_1^\gamma M_0^{\gamma*}}{a\sum_{\rm pol}|M_0^\gamma|^2}-h_V,\qquad
h_V=\frac{2\,\mathrm{Re}\,M_1M_0^*}{a\,|M_0|^2},
\end{equation}
and
\begin{equation}
\Delta_{RV}(E)=\frac1{4EPK_m^2}\int_0^{K_m}\!dk\,(K_m-k)^2\,\Omega(E,k),\qquad
\Omega(E,k)=\int_{s_-}^{s_+}\!ds\,\delta\,T_0,\qquad s=2p\cdot k,\ s_\pm=2k(E\pm P).
\end{equation}
Here $T_0$ is the polarization-summed tree trace with one photon. The factor $(K_m-k)^2/K_m^2$ is the static neutrino phase space. The wave-function and current
counterterms cancel between $M_1^\gamma$ and $M_1$. $\delta$ is UV and IR finite and vanishes like $k\ln k$ for $k\to0$ (Low's theorem).

\paragraph{Diagrams and masters.} The loop photon is either on the electron line (self-energy of the intermediate electron, vertex
correction of the emission vertex: $A_1$, $A_2$), or exchanged between the electron and the static proton (photon emitted before or after the
exchange: $B_1$, $B_2$). The non-radiative vertex $V$ is subtracted in $h_V$. Passarino--Veltman reduction (only rank-1 tensors survive) gives
\begin{equation}
\delta\,T_0=\mathrm{Re}(T_{A_1}+T_{A_2})+c_{\rm box}\Bigl(\mathrm{Box}-\frac{T_p}s\Bigr)+c_{T_q}T_q+c_{T_{pq}}T_{pq}+c_{C_0}C_0+c_{B_p}B_p+c_{B_q}B_q+c_{S_p}S_p+c_{S_q}S_q .
\label{eq:deltaRV}
\end{equation}
The real parts of the masters are understood. Define $D_0=l^2$, $D_p=(l-p)^2-1$, $D_q=(l-q)^2-1$, $D_S=v\cdot l$, $q=p+k$, $q^0=E+k$, $P_q=\sqrt{q_0^2-1}$,
$|\vec q|=\sqrt{q_0^2-1-s}$, $\varphi_E=\operatorname{arccosh}E$ and $\varphi_q=\operatorname{arccosh}q^0$. With every integral $\int d^dl/(i\pi^{d/2})$ normalized to $\Gamma(1+\epsilon)$, the finite real parts are
\begin{align}
B_p&=2,\quad B_q=2-\frac s{1+s}\ln s,\quad S_p=4E-2PL_\beta,\quad S_q=4q^0-4P_q\varphi_q,\nonumber\\
T_{pq}&=\frac{\varphi_q^2-\varphi_E^2}{k},\qquad C_0=\frac{\ln s\ln(1+s)+\mathrm{Li}_2(-s)}{s},\qquad T_q=-2\,\mathcal H(q^0,|\vec q|,P_q),\nonumber\\
\mathcal H(a,R,w_1)&=\frac1{2R}\Bigl[2\mathrm{Li}_2(\alpha)-\frac{\pi^2}3+\frac12\ln^2\alpha-\ln\alpha\ln\tau_1+\mathrm{Re}\,\mathrm{Li}_2\Bigl(\frac{\tau_1}{\alpha}\Bigr)-\mathrm{Li}_2(\alpha\tau_1)\Bigr],\quad
\alpha=\frac{a-R}{a+R},\ \tau_1=\frac{w_1-R}{w_1+R},\nonumber\\
\mathrm{Box}-\frac{T_p}s&=\frac1s\Bigl[-\mathcal X-\frac{L_\beta}P\ln\frac{4P^2}{s}+\frac1P\Bigl(2\mathrm{Li}_2(\alpha_E)-\frac{\pi^2}3+\frac12L_\beta^2\Bigr)\Bigr],\qquad\alpha_E=\frac{E-P}{E+P},\\
\mathcal X&=\frac2P\Bigl[\mathrm{Li}_2\Bigl(1-\frac{\xi_q}{\xi_+}\Bigr)+F\Bigl(\frac{\xi_-}{\xi_q}\Bigr)-F\Bigl(\frac{\xi_-}{\xi_+}\Bigr)\Bigr]\nonumber\\&\quad+\frac{L_\beta}P\ln\frac{(\xi_q-\xi_-)\xi_+}{\xi_q(\xi_+-\xi_-)},\nonumber\\
F(t)&=-\ln\xi_-\ln t+\tfrac12\ln^2t+\ln\xi_-\ln(1-t)+\mathrm{Li}_2(1-t),\qquad\xi_\pm=E\pm P,\ \xi_q=q^0+P_q.\nonumber
\end{align}
The coefficients $c_i$, $T_0$ and $\mathrm{Re}(T_{A_1}+T_{A_2})=a_0+a_s\ln s+a_{\ell\ell}\ln s\ln(1+s)+a_{\rm Li}\mathrm{Li}_2(-s)$ are rational functions of $(E,k,s)$.
They are listed below ($Q^2\equiv|\vec q|^2=q_0^2-1-s$; $T_0$ without the factor $1+3\lambda^2$). The UV poles cancel identically ($c_{B_p}+c_{B_q}+2Ec_{S_p}+2q^0c_{S_q}=0$). The IR-divergent $T_p$ enters only through
the finite combination $\mathrm{Box}-T_p/s$.
{\small
\begin{align*}
T_0&=- \frac{2 \left(- 4 E^{2} k s - 4 E k^{2} s + 4 E k^{2} + E s^{2} - 2 k^{3} s + 4 k^{3} + k s^{2}\right)}{k^{2} s^{2}}\\
c_{\rm box}&=\frac{- 8 E^{3} s - 8 E^{2} k s + 16 E^{2} k - 4 E k^{2} s + 16 E k^{2} + E s^{2}}{k s}\\
c_{T_q}&=\frac{1}{2 Q^{2} k s^{2}}\Bigl[- 16 E^{5} s - 32 E^{2} k - 64 E k^{2} + 32 k^{5} - 24 k^{3} s - 32 k^{3}\\
&\qquad- 11 E s^{3} - 10 E s^{2} - 8 k^{5} s + 14 k^{3} s^{2} - 7 k s^{3} - 8 k s^{2}\\
&\qquad+32 E^{4} k + 128 E^{3} k^{2} + 26 E^{3} s^{2} + 16 E^{3} s + 192 E^{2} k^{3} + 128 E k^{4}\\
&\qquad- 64 E^{4} k s - 104 E^{3} k^{2} s - 88 E^{2} k^{3} s - 40 E k^{4} s + 50 E k^{2} s^{2} - 40 E k^{2} s\\
&\qquad+62 E^{2} k s^{2}\Bigr]\\
c_{T_{pq}}&=\frac{- 8 E^{2} - 8 E k - 4 k^{2} + s}{s}\\
c_{C_0}&=\frac{8 E^{2} + 8 E k + 4 k^{2} - s}{k}\\
c_{B_p}&=\frac{1}{k s^{2} \left(E - 1\right) \left(E + 1\right)}\Bigl[- 8 E^{3} k s + 8 E^{2} k^{2} + 2 E^{2} s^{2} + 8 k^{2} s - 8 k^{2} - 3 s^{2}\\
&\qquad- 8 E^{2} k^{2} s + 10 E k s\Bigr]\\
c_{B_q}&=\frac{1}{Q^{2} k s^{2}}\Bigl[- 16 E k^{3} - 8 k^{4} - 10 k^{2} s^{2} + 8 k^{2} + 3 s^{3} + 3 s^{2}\\
&\qquad- 8 E^{2} k^{2} - 2 E^{2} s^{2} - 16 E k s^{2} - 10 E k s + 8 k^{4} s + 2 k^{2} s\\
&\qquad+8 E^{3} k s + 24 E^{2} k^{2} s + 24 E k^{3} s\Bigr]\\
c_{S_p}&=\frac{1}{2 E k^{2} s^{2} \left(E - 1\right) \left(E + 1\right)}\Bigl[- 8 E^{3} k^{2} - 2 E^{3} s^{2} - 8 E^{2} k^{3} + 2 E s^{2} - 4 k^{3} s + 8 k^{3}\\
&\qquad+4 E^{2} k^{3} s - E^{2} k s^{2} - 8 E^{2} k s - 10 E k^{2} s + 8 E k^{2} + 2 k s^{2}\\
&\qquad+8 E^{4} k s + 8 E^{3} k^{2} s\Bigr]\\
c_{S_q}&=\frac{1}{2 Q^{2} k^{2} s^{2}}\Bigl[- 4 k^{4} s + 8 k^{4} - 6 k^{2} s - 8 k^{2} - 2 s^{3} - 2 s^{2}\\
&\qquad- 16 E^{2} k^{2} s + 8 E^{2} k^{2} + 2 E^{2} s^{2} - 12 E k^{3} s + 16 E k^{3} + 5 k^{2} s^{2}\\
&\qquad- 8 E^{3} k s + 11 E k s^{2} + 8 E k s\Bigr]
\end{align*}
\noindent with $\mathrm{Re}(T_{A_1}+T_{A_2})=a_0+a_s\ln s+a_1\ln(1+s)+a_{\ell\ell}\ln s\ln(1+s)+a_{\rm Li}\,\mathrm{Li}_2(-s)$ and
\begin{align*}
a_{\rm Li}&=\frac{- 8 E k s - 8 k^{2} s + 8 k^{2} + 2 s^{2}}{k s^{3}}\\
a_{\ell\ell}&=\frac{- 8 E k s - 8 k^{2} s + 8 k^{2} + 2 s^{2}}{k s^{3}}\\
a_{1}&=0\\
a_{s}&=\frac{1}{k s^{2} \left(s + 1\right)^{2}}\Bigl[8 k^{2} s^{2} - 12 k^{2} s - 8 k^{2} - 2 s^{4} - 5 s^{3} - 2 s^{2}\\
&\qquad+8 E k s^{3} + 20 E k s^{2} + 8 E k s + 8 k^{2} s^{3}\Bigr]\\
a_{0}&=\frac{- 4 E k s^{2} - 8 E k s - 4 k^{2} s^{2} + 8 k^{2} + s^{3} + 2 s^{2}}{k s^{2} \left(s + 1\right)}
\end{align*}
}

\paragraph{Fixed photon energy: closed form.} The angular integral $\Omega(E,k)$ is elementary for every term of Eq.~(\ref{eq:deltaRV}) except
$c_{T_q}T_q$, giving $\ln$, $\mathrm{Li}_2(-s_\pm)$, $\mathrm{Li}_2(s_\pm/s_0)$ and $\mathrm{Li}_3(-s_\pm)$ with $s_0=P_q^2$. The $T_q$ term collapses to
\begin{equation}
\Omega_{T_q}=-2\int_{P_q}^\infty\!dw\,\ln\Bigl|\frac{q^0-w}{q^0+w}\Bigr|\Bigl\{c_aL+c_b\Bigl[\frac{L-L_\beta}{\omega^2}+\frac{P/k}{\omega}\Bigr]+c_c\frac{L-L_\beta}{\omega}+c_d\frac{L-\ell_1}{w^2}\Bigr\},
\end{equation}
with $\omega=w^2-P_q^2$, $L=\ln\frac{w^2-(P-k)^2}{w^2-(P+k)^2}$, $\ell_1=2\ln\frac{|P-k|}{P+k}$, and
$c_a=\frac{11E+7k}{2k}$, $c_b=16q_0^2$, $c_c=\frac{4q^0(2E^2+2Ek+k^2)}{k}$, $c_d=\frac{(E-k)(1+k^2-E^2)}{2k}$. In $\tau=(w-P_q)/(w+P_q)$ all
logarithms become $\ln|1-a\tau|$, and
\begin{equation}
\Omega_{T_q}=-2\sum_{i=a}^dc_i\bigl[F_i(\zeta,R_-)-F_i(\zeta,R_+)\bigr]-c_b\frac{P}{kP_q}\bigl[\mathrm{Li}_2(\zeta^{-2})-\mathrm{Re}\,\mathrm{Li}_2(\zeta^2)\bigr],
\qquad \zeta=q^0+P_q,\ R_\pm=\frac{P_q-(P\pm k)}{P_q+(P\pm k)} .
\end{equation}
The functions $F_{a,b,c,d}(\zeta,R)$ contain only $\mathrm{Li}_2$, logarithms, and the single weight-3 function
\begin{equation}
S(A,B)=\int_0^1\frac{dt}{t}\ln|1-At|\ln|1-Bt|=\tfrac12\bigl[K(A)+K(B)-M(A,B)\bigr],
\end{equation}
where $K(x)=\ln x\ln^2(1-x)+2\ln(1-x)\mathrm{Li}_2(1-x)-2\mathrm{Li}_3(1-x)+2\zeta_3$, $M=\Phi_1(U)-\Phi_{B/A}(U)+2\zeta_3-2\mathrm{Li}_3(B/A)$, $U=\frac{1-A}{1-B}$,
$\Phi_c(u)=\ln^2u\ln(1-cu)+2\ln u\,\mathrm{Li}_2(cu)-2\mathrm{Li}_3(cu)$ (real parts). The $F_i$ are listed in the supplementary file
\texttt{RV\_tq\_blocks.tex}. The four blocks cancel like $k^{-2}$ at small $k$, so numerically one needs $\approx2|\log_{10}k|$ extra digits.
The closed form agrees with direct quadrature of the angular integral to 28--30 digits.

\paragraph{Photon-energy integral.} The $k$ integral raises the weight to four, with the additional square-root letter $P_q$. For the numbers in
the tables it is done by Gauss quadrature in $t$, with $k=K_mt^2$; 16 points converge to $10^{-5}$ relative. As an independent check, $\Delta_{RV}(E)$ was obtained exactly at ten
electron energies with rational $\xi=E+P$, as sums of 1600--2600 Goncharov polylogarithms of weight $\le4$ in the variable $\zeta$. The two
evaluations agree to $\lesssim10^{-5}$ (24 points in $k$).

\subsection{Double-real remainder $D$}
With $x_i=p\cdot k_i=k_i(E-Pc_i)$ and $y=k_1\cdot k_2=k_1k_2(1-c_1c_2)-k_1k_2\sqrt{(1-c_1^2)(1-c_2^2)}\cos\phi$,
\begin{equation}
D(E)=\frac12\int\!dk_1dk_2\,dc_1dc_2\,\frac{k_1k_2}{2\pi}\Bigl(\frac{K_m-k_1-k_2}{K_m}\Bigr)^2\theta(K_m-k_1-k_2)\int_0^{2\pi}\!\!d\phi\,
\Bigl[\frac{S_2}{2E}-\frac{S_1(k_1)S_1(k_2)}{(2E)^2}\Bigr] .
\end{equation}
$S_2$ ($S_1$) is the polarization-summed trace for two (one) photons emitted from the electron line, in the gauge above. It is given by the FORM file
\texttt{dr/rr.frm} (Levi-Civita terms vanish after the $\phi$ integral). The second term is the independent-emission product under the
\emph{two}-photon phase space. The only $\phi$-dependent denominator is $(p+k_1+k_2)^2-1=2(x_1+x_2+y)$, so the azimuthal integral is closed form,
\begin{equation}
\langle u^{-1}\rangle=(a^2-B^2)^{-1/2},\quad\langle u^{-2}\rangle=a(a^2-B^2)^{-3/2},\quad\langle u^{-3}\rangle=\frac{2a^2+B^2}{2(a^2-B^2)^{5/2}},
\end{equation}
with $u=a-B\cos\phi$, $a=x_1+x_2+k_1k_2(1-c_1c_2)$ and $B=k_1k_2\sqrt{(1-c_1^2)(1-c_2^2)}$ (\texttt{dr/dphi.py}). The remaining four integrals are done by Gauss
quadrature with $k_1=K_mu^2$, $k_2=(K_m-k_1)v^2$ and a logarithmic map of each photon angle, $E-Pc=(E-P)\bigl(\frac{E+P}{E-P}\bigr)^{t}$, which resolves the
collinear peak at large $E$ (\texttt{dr/dnum.py}, function \texttt{Dlog}). With $32^4$ points the result is converged to $\sim10^{-5}$ relative.

As an independent check, $D(E)$ was obtained exactly at ten electron energies as sums of Goncharov polylogarithms of weight $\le4$
($\sim2\times10^4$ per energy, plus a one-dimensional double-soft corner constant). The two angular integrations are done analytically, and the
photon-energy integrals use rationalizing variables $E+k_1=\frac12(u+u^{-1})$, $E+k_1+k_2=\frac12(z+z^{-1})$. The quadrature agrees with the exact values to about $10^{-5}$ relative, consistent with its estimated accuracy; it lies systematically above them, by $0.1$--$4\times10^{-6}$ in units of $(\alpha/2\pi)^2$.

\subsection{Phase-space mismatch $\Delta_{PS}$}
$\frac12(H_R^{(1)})^2$ is the product of independent one-photon integrals with weights $(1-k_i/K_m)^2$, while the subtraction in $D$ uses the joint weight
$((K_m-k_1-k_2)/K_m)^2$ with $k_1+k_2<K_m$. Using the exact one-photon spectrum $\rho(k)=u/k+(2L_\beta/P-4/E)+L_\beta k/(PE)$, the difference is
\begin{equation}
\Delta_{PS}=\Bigl(\frac58-\frac{\pi^2}{12}\Bigr)u^2-\frac{uv}9-\frac{7uw}{144}-\frac{v^2}{72}-\frac{vw}{90}-\frac{w^2}{480},\qquad
u=\frac{2L_\beta}{\beta}-4,\ v=\Bigl(\frac{2L_\beta}P-\frac4E\Bigr)K_m,\ w=\frac{L_\beta K_m^2}{PE}.
\end{equation}

\subsection{Positron emitters}
In the static limit a spectator charge $Z$ does not radiate. The $Z^0$ part of a positron decay $(Z+1)\to Z\,e^+\nu$ is therefore the process with an
incoming static charge and an outgoing positron. All photon couplings change sign together, so $H_R^{(1)}$, $D$ and $\Delta_{PS}$ are unchanged.
In the loops the static propagator becomes $1/(v\cdot l-i0)$. The difference, $2\pi i\,\delta(v\cdot l)$, has a real part only in the IR-singular
(Coulomb) region:
\begin{itemize}
\item it shifts $T_p$ by $\pi^2/P$, which reproduces exactly $H_V^{(1)}\to H_V^{(1)}-2\pi^2/\beta$;
\item it leaves $S_{p,q}$, $T_q$ and $T_{pq}$ unchanged (their $\delta$-parts are purely imaginary);
\item it leaves the combination $\mathrm{Box}-T_p/s$ unchanged.
\end{itemize}
In the real-virtual remainder $\delta(E,k,\theta)$, Eq.~(\ref{eq:deltaRV}), $T_p$ appears only in this combination: its coefficients from $B_1$, $B_2$ and the subtracted $h_V$ add up to $-c_{\rm box}/s$. Hence $\Delta_{RV}$ is identical for $e^\pm$. A direct numerical evaluation of the $\delta(v\cdot l)$ parts confirms this
to $10^{-12}$ (\texttt{analytic/crossing\_check.py}). For positron emitters $H_V^{(2)}$ is the form factor at the real point $w$ (no continuation).

\subsection{Real $e^+e^-$ pair emission}
\label{app:pairs}
For $W_0>3$ the final state $e\,e^+e^-\nu$ (plus the static source) is open. At $O(\alpha^2)$ it must be included for the rate to have a
finite $m\to0$ limit. The pair is produced by a virtual photon $q=p_2+p_3$ emitted from the electron line.
As for real photons, we use the propagator numerator $P_{\mu\nu}(q)=-g_{\mu\nu}+(q_\mu v_\nu+v_\mu q_\nu)/v\!\cdot\! q-q_\mu q_\nu/(v\!\cdot\! q)^2$.
This is allowed because the pair current is conserved, and since $v\cdot P=0$ the static line does not couple.
With two identical electrons the amplitude is $A-B$, where $B$ is $A$ with $p_1\leftrightarrow p_2$. Including the factor $1/2!$, the rate is
\begin{equation}
\Gamma_{e^+e^-}\propto\int d\Phi\,\Bigl[\,|A|^2-\mathrm{Re}\,(AB^*)\Bigr].
\end{equation}

\paragraph{Direct term.} $|A|^2$ contains a closed electron loop and is $\propto n_e$. With $p_1$ the electron whose energy $E$ is the spectrum variable,
$K_m=W_0-E$, and units $(\alpha/2\pi)^2$, the Kroll--Wada factorization gives
\begin{equation}
X_{\rm dir}(E)=\int_{4}^{K_m^2}\!\frac{ds}{s}\,\frac23\Bigl(1+\frac2s\Bigr)\sqrt{1-\frac4s}\;\int_{\sqrt s}^{K_m}\!dk^0\,|\vec k|\int_{-1}^{1}\!dc\;
\frac{T_{\gamma^*}(s)}{2E}\Bigl(\frac{K_m-k^0}{K_m}\Bigr)^2 .
\end{equation}
Here $T_{\gamma^*}$ is the polarization-summed one-photon tree trace $T_0$ for a photon of mass $\sqrt s$, i.e.\ with denominator $2p\cdot k+s$ and the sum $P_{\mu\nu}(k)$.
At $s\to0$ it reproduces the one-photon spectrum $\rho(k)$ exactly (\texttt{pair/pair.py}).

\paragraph{Exchange term.} In $\mathrm{Re}(AB^*)$ the two electron lines join into a single fermion line, so it carries no factor $n_e$ and belongs to the
photon-only sector. It is the QED analogue of the $Q Q\bar Q$ cut $G^{(3Q)}$ in the $C_F(C_F-C_A/2)$ part of the two-loop massive quark jet function~\cite{Hoang:2019fze}. 
\begin{equation}
T_{\rm ex}=\mathrm{Tr}\Bigl[(\slashed p_1+1)\,\gamma^\mu\frac{\slashed P+1}{P^2-1}\,\mathrm{Mid}\,\frac{\slashed P+1}{P^2-1}\gamma^\nu\,(\slashed p_2+1)\,
\frac{P_{\mu\mu'}(q)}{q^2}\gamma^{\mu'}(\slashed p_3-1)\frac{P_{\nu\nu'}(q')}{q'^2}\gamma^{\nu'}\Bigr],
\end{equation}
where $P=p_1+p_2+p_3$, $q=p_2+p_3$, $q'=p_1+p_3$ and $\mathrm{Mid}=\frac{1+\gamma_5}{2}\gamma^0$ (neutrino-averaged), the exchange contribution is
\begin{equation}
X_{\rm ex}(E)=-\frac1{2\pi}\int dE_2\,dE_3\,dc_2\,dc_3\,d\phi\;|\vec p_2||\vec p_3|\,\frac{\mathrm{Re}\,T_{\rm ex}}{2E}\Bigl(\frac{K_m-E_2-E_3}{K_m}\Bigr)^2,\qquad E_2+E_3\le K_m .
\end{equation}
We evaluate it by quasi-Monte Carlo, with logarithmic maps of the polar angles (\texttt{interf/interf.py}). The same code with $T_{\rm dir}$
reproduces the Kroll--Wada result, e.g.\ $0.05753$ vs.\ $0.05772$ at $(E,W_0)=(3,8)$. For the nuclei of Table~\ref{tab:nuc} the exchange term is
15--25\% of the direct term.

\paragraph{Finiteness check.} At fixed $\mu=\mu_{UV}=2W_0$ and on-shell $\alpha$, the strict $O(\alpha^2)$ coefficient must behave as
$-\frac23 L_m K_1+\mathrm{const}$ for $m/W_0\to0$ (finite in $\bar\alpha$). Here $L_m=\ln(m^2/\mu^2)$ and $K_1\to\Delta^{(1,0)}/a$, i.e.\ its slope
in $\ln W_0$ is $\frac43 c_1$ with $c_1=\frac{217}{20}-\frac{4\pi^2}{3}$~\cite{Cao:2025zxs}. We checked: 
\begin{itemize}
\item 
the electron-loop sector, the $n_e$ part of $H_V^{(2)}$ plus the direct pair term, was evaluated for
$W_0=1280$--$5.2\times10^{6}$, with the pair integral computed to a relative precision of $\sim10^{-9}$
(\texttt{pair/nl\_slope.py}). This is needed because the two pieces cancel to about $2\%$ at large $W_0$.
A fit $a+s\ln W_0+b/W_0$ gives $s=-3.079303\pm0.000003$, against $\frac43c_1=-3.079297$; equivalently, it
reproduces $c_1$ to $2\times10^{-6}$.
\item the photon-only sector without the exchange term has slope $-1.427\pm0.003$, independent of the neutrino weight ($E_\nu^2$ or $E_\nu^1$);
\item the exchange term cancels it: photon-only plus exchange is $22.43(4)$, $22.42(10)$, $22.23(19)$, $22.17(26)$ at
$W_0=80,\,320,\,1280,\,5120$, where the uncertainties are those of the quasi-Monte Carlo evaluation of the exchange term
(\texttt{interf/interf\_scan.py}). A fit in $\ln W_0$ gives a slope of $-0.05\pm0.04$, consistent with zero, compared with
$-1.427$ without the exchange term.
\end{itemize}
Without the exchange term the result would have an uncanceled $\ln(W_0/m)$. Its numerical effect on Table~\ref{tab:nuc} is at most $0.0003\times10^{-3}$.

\subsection{Reference values}
\begin{center}\small
\begin{tabular}{cccc}\toprule
$\xi=E+P$ & $E/m$ & $\Delta_{RV}(E)$ & $D(E)$\\\midrule
11/10 & 1.004545 & $+0.14146417$ & 0.081015187\\
6/5 & 1.016667 & $+0.05056442$ & 0.103047460\\
3/2 & 1.083333 & $-0.41810972$ & 0.205692256\\
2 & 1.250000 & $-1.35497113$ & 0.358372012\\
5/2 & 1.450000 & $-2.06444306$ & 0.417767635\\
3 & 1.666667 & $-2.39611522$ & 0.396843965\\
7/2 & 1.892857 & $-2.33774570$ & 0.323049894\\
4 & 2.125000 & $-1.90199612$ & 0.217446587\\
9/2 & 2.361111 & $-1.05703515$ & 0.093915503\\
24/5 & 2.504167 & $-0.23897882$ & 0.014994190\\\bottomrule
\end{tabular}\end{center}
These are the exact values for neutron decay ($\Delta=1.29333236/0.51099895$), obtained as sums of Goncharov polylogarithms evaluated numerically with GiNaC~\cite{Bauer:2000cp,Vollinga:2004sn}. Averaged over the static spectrum with the soft
factor (unresummed), the remainders contribute $\langle a^2\Delta_{RV}\rangle=-0.00255$, $\langle a^2D\rangle=+0.00046$ and $\langle a^2\Delta_{PS}\rangle=-0.00078$, all in units of $10^{-3}$.
A tabulation on a dense grid is provided in the supplementary data (App.~\ref{app:spectrum}).

\section{Radiative correction to the electron spectrum: numerical recipe}
\label{app:spectrum}

\paragraph{Spectrum.}  (For a review of the standard description of allowed $\beta$ spectra see Ref.~\cite{Hayen:2017pwg}.) In the static limit, the electron spectrum of neutron decay, inclusive over photons, is
\begin{equation}
\frac{d\Gamma}{dE}=\frac{G_F^2|V_{ud}|^2(1+3\lambda^2)\,m^5}{2\pi^3}\,|C(\mu_{UV})|^2\;pE(\Delta-E)^2\;R(E),
\label{eq:spec}
\end{equation}
in units $m=m_e=1$, with $1\le E\le\Delta$ and $p=\sqrt{E^2-1}$.
$|C(\mu_{UV})|^2$ contains all short-distance (electroweak, hadronic) corrections. Recoil, finite-size and other corrections not discussed here multiply Eq.~(\ref{eq:spec}).
The QED factor through $O(\alpha^2)$ is
\begin{align}
R(E)&=S(E)\Bigl[1+aH^{(1)}(E)+a^2H^{(2)}(E)\Bigr]&&\text{(unresummed)},\label{eq:Runres}\\
R(E)&=S(E)\,\tilde F(E)\Bigl[1+a\tilde H^{(1)}(E)+a^2\tilde H^{(2)}(E)\Bigr]&&\text{(with Fermi function)},\label{eq:RF}
\end{align}
where $a=\alpha/2\pi$ and $\alpha$ is on shell. The other ingredients are:
\begin{itemize}
\item $S$ and $H_R^{(1)}$ from App.~\ref{app:real};
\item $H^{(1)}=H_V^{(1)}+H_R^{(1)}$, with $H^{(2)}$ as in Eq.~(\ref{eq:H2decomp});
\item $H_V^{(1,2)}$ from Eqs.~(\ref{eq:HV1})--(\ref{eq:H2V}) at $w=E$, converted from $\mu_{UV}=\mu$ to $\mu_{UV}$ by
\begin{align}
H_V^{(1)}&\to H_V^{(1)}+3\ell_{UV},\qquad\ell_{UV}=\ln\frac{\mu_{UV}}\mu,\nonumber\\
H_V^{(2)}&\to H_V^{(2)}+3\ell_{UV}H_V^{(1)}+\tfrac92\ell_{UV}^2+\Bigl(\tfrac{4\pi^2}3-\tfrac{35}{12}\Bigr)\ell_{UV}+2\bigl(\ln^2\mu_{UV}-\ln^2\mu\bigr);
\end{align}
\item $\tilde F$, Eq.~(22) of Ref.~\cite{Cao:2025zxs} 
 with $Z\to-1$, at the scale $\mu$;
\item $\tilde H^{(1)}=H^{(1)}-f_1$ and $\tilde H^{(2)}=H^{(2)}-f_1H^{(1)}+f_1^2-g_2$, with $f_1=2\pi^2/\beta$ and $g_2=\frac{4\pi^4}{3\beta^2}+4\pi^2\bigl(\frac{11}4-\ln\frac{2p}\mu\bigr)$.
\end{itemize}
$R(E)$ is independent of $\mu$ and of the photon cutoff $\epsilon_\gamma$ up to $O(\alpha^3)$. Its $\mu_{UV}$ dependence cancels against $|C(\mu_{UV})|^2$.
Equations~(\ref{eq:Runres}) and (\ref{eq:RF}) differ at $O(\alpha^3)$. 

\paragraph{Ingredients.} All ingredients are closed-form functions except $\Delta_{RV}(E)$ and $D(E)$:
\begin{itemize}
\item $H_V^{(1,2)}$: classical polylogarithms of weight $\le4$ in $x=E-p$;
\item $H_R^{(1)}$, $S$, $\Delta_{PS}$: elementary functions and $\mathrm{Li}_2$.
\end{itemize}
$\Delta_{RV}(E)$ and $D(E)$ are smooth, $\mu$- and $\epsilon_\gamma$-independent functions of $E$. They are tabulated in \texttt{nbd\_real\_grid.csv} on the grid
$p_i=i\,p_{\max}/100$, $i=0,\dots,100$. Both vanish linearly (up to logarithms) at the endpoint:
$\Delta_{RV}\simeq K_m(c_0+c_1\ln K_m)$, $D\simeq c_0'K_m$, with $K_m=\Delta-E$. We therefore interpolate $f/K_m$ in $\ln K_m$ with local cubic polynomials.
This reproduces the exact node values of App.~\ref{app:real} to $\le7\times10^{-6}$ for $\Delta_{RV}$ and $\le4\times10^{-6}$ for $D$. That is the
accuracy of the tabulation itself, and far below any experimental resolution: $a^2\times10^{-5}\approx10^{-10}$.

\paragraph{Script.} The supplementary script \texttt{numerics/spectrum/spectrum.py} (Python~3 with \texttt{mpmath}; no other dependency) implements
Eqs.~(\ref{eq:spec})--(\ref{eq:RF}) with the inputs $\alpha^{-1}=137.035999177$, $m_e=0.51099895$\,MeV and $m_n-m_p=1.29333236$\,MeV:
\begin{verbatim}
python3 spectrum.py                     # Table-1 check, writes spectrum_R.csv
python3 spectrum.py --E 1.8             # all ingredients at E = 1.8 m_e
python3 spectrum.py --mu 2 --eps 0.5    # mu = 2 m_e, eps_gamma = 0.5 Delta
\end{verbatim}
The options \texttt{--mu} and \texttt{--eps} set the scale $\mu$ (in units of $m_e$; default $\sqrt{m_e\Delta}$) and the photon cutoff $\epsilon_\gamma$
(in units of $\Delta$; default $\Delta$). $R(E)$ changes only at $O(\alpha^3)$, which provides a check.
The function \texttt{R(E)} returns $R$ at $O(\alpha)$ and $O(\alpha^2)$ in both forms, together with all ingredients ($S$, $H_V^{(1,2)}$, $H_R^{(1)}$, $\Delta_{RV}$, $D$,
$\Delta_{PS}$, $\tilde F$). The spectrum average
\begin{equation}
\delta_{R,\rm static}=\frac{\int_1^\Delta dE\,pE(\Delta-E)^2\,[R(E)-1]}{\int_1^\Delta dE\,pE(\Delta-E)^2}
\end{equation}
(40-point Gauss--Legendre in $p$) gives, at the central scales,
\begin{center}
\begin{tabular}{lcc}\toprule
& $O(\alpha)$ & $O(\alpha^2)$\\\midrule
unresummed, Eq.~(\ref{eq:Runres}) & 28.80381 & 29.19922\\
with $\tilde F$, Eq.~(\ref{eq:RF}) & 29.23011 & 29.19827\\\bottomrule
\end{tabular}
\end{center}
in units of $10^{-3}$, reproducing the first and last rows of Table~\ref{tab:table1_fullF}. Running with $\mu=m/2$ and $\mu=2\Delta$ reproduces the $\mu$ spread $0.0008$
(unresummed) and $0.0006$ (with $\tilde F$). The file \texttt{spectrum\_R.csv} lists $R(E)$ on 199 equally spaced momenta. Linear interpolation in it is accurate to $3\times10^{-4}$ for
$E\ge1.01\,m_e$. For higher accuracy, or near threshold, where $R$ grows like $1/\beta$, evaluate \texttt{R(E)} directly (about 50\,ms per energy).

\paragraph{Nuclear $\beta^\mp$ decays ($Z^0$ part).} Equation~(\ref{eq:Runres}) applies to the $Z$-independent part of nuclear decays with
$\Delta\to W_0$. For $\beta^+$ make the replacements of App.~\ref{app:real}:
\begin{itemize}
\item $H_V^{(1)}\to H_V^{(1)}-2\pi^2/\beta$;
\item $H_V^{(2)}\to$ the form factor at the real point $w$ (\texttt{numerics/tables/hv\_cross.py});
\item $\Delta_{RV}$, $D$, $\Delta_{PS}$ and $H_R^{(1)}$ unchanged.
\end{itemize}
$\Delta_{RV}$ and $D$ depend on $W_0$ through $K_m=W_0-E$ and must be recomputed for each $W_0$ (\texttt{numerics/tables/nuclear\_real.py}, a few CPU-minutes per nucleus).  
Table~\ref{tab:nuc} uses $\mu=\mu_{UV}=\mu_L=2W_0$, $\epsilon_\gamma=W_0$ and the static phase space $pE(W_0-E)^2$ weighted with $\tilde F$ (\texttt{numerics/tables/table2.py}).

\section{Resummation schemes}
\label{app:resummation}

In nuclear beta decay a Fermi function is factored off from the rest of the radiative corrections. This resums a series in $Z\alpha/\beta$, and improves the convergence of perturbation theory. 
In neutron decay the same scheme (with a Fermi function) is traditionally used.  However, here $Z=1$ and $\beta \sim O(1)$, 
and there is not an obvious small expansion parameter governing a
systematic resummation.  More recently, an argument based on tracking $\pi$-enhancements from RG-evolution was put forth that reproduces the leading $\pi \alpha/\beta$ term in the Fermi function \cite{VanderGriend:2025mdc,Cao:2025lrw}, together with a series of higher-order corrections,  
and can be justified without appealing to the non-relativistic portion of the spectrum. 

In the main text we have presented our results both 
at fixed orders in the $\alpha$ expansion 
(``without Fermi function''),
and at fixed orders after extracting a traditional Fermi function prefactor (``with Fermi function''). Strictly speaking, one should treat the $\beta \sim O(\alpha)$ and $\beta \sim O(1)$ regions differently, resumming with the Fermi function in the former, and resumming $\pi^2$ enhancements in the latter.
At $O(\alpha^3)$ accuracy, however, hydrogenic bound states $n\rightarrow \!~^{1}{\rm H}_{n\ell m} \bar{\nu}_e$ must also be included in the calculation of the lifetime \cite{Cao:2025lrw}. Since $O(\alpha^3)$ is beyond our desired accuracy we refrain from a more detailed discussion.  As is clear from \cref{tab:table1_fullF}, with the complete two-loop results in place, 
expanding with or without Fermi function prefactor 
gives equivalent values for $\delta_{R,~{\rm static}}$ within experimental precision.

\end{fmffile}

\bibliography{references.bib}

\end{document}